\documentclass[english,aps,prc,nofootinbib,superscriptaddress,showpacs,twocolumn,letter]{revtex4}
\usepackage{amsmath}
\usepackage{CJK}
\usepackage[    bookmarks,
                 bookmarksopen = true,
                 bookmarksnumbered = true,
                 linktocpage,
                 colorlinks = true,
                 linkcolor = blue,
                 urlcolor  = red,
                 citecolor = blue,
                 anchorcolor = green,
                 hyperindex = true,
                 hyperfigures]
		{hyperref}
\usepackage[utf8]{inputenc}
\usepackage{graphicx}
\usepackage{esint}
\usepackage{xcolor}
\usepackage{hyperref}

\usepackage{graphicx}  
\usepackage{float}  
\usepackage{subfigure}

\makeatletter

\@ifundefined{textcolor}{}
{%
 \definecolor{BLACK}{gray}{0}
 \definecolor{WHITE}{gray}{1}
 \definecolor{RED}{rgb}{1,0,0}
 \definecolor{GREEN}{rgb}{0,1,0}
 \definecolor{BLUE}{rgb}{0,0,1}
 \definecolor{CYAN}{cmyk}{1,0,0,0}
 \definecolor{MAGENTA}{cmyk}{0,1,0,0}
 \definecolor{YELLOW}{cmyk}{0,0,1,0}
}
\makeatother

\usepackage{babel}

\newcommand{\eg}{e.g.,~}

\begin{document}
\begin{CJK*}{GB}{}
\preprint{This line only printed with preprint option}
\title{Directed flow of charged particles within idealized viscous hydrodynamics at energies available at the BNL Relativistic Heavy Ion Collider and at the CERN Large Hadron Collider}

\author{Ze-Fang Jiang}
\email{jiangzf@mails.ccnu.edu.cn}
\affiliation{Department of Physics and Electronic-Information Engineering, Hubei Engineering University, Xiaogan 432000, China}
\affiliation{Institute of Particle Physics, Central China Normal University, Wuhan 430079, China}
\affiliation{Key Laboratory of Quark and Lepton Physics, Ministry of Education (MOE), Wuhan, 430079, China}

\author{C. B. Yang}
\email{cbyang@mail.ccnu.edu.cn}
\affiliation{Institute of Particle Physics, Central China Normal University, Wuhan 430079, China}
\affiliation{Key Laboratory of Quark and Lepton Physics, Ministry of Education (MOE), Wuhan, 430079, China}

\author{Qi Peng}
\affiliation{Department of Physics and Electronic-Information Engineering, Hubei Engineering University, Xiaogan 432000, China}

\selectlanguage{english}%

\begin{abstract}
Following the Boz$\dot{\textrm{e}}$k-Wyskiel parametrization tilted initial condition,
an alternative way to construct a longitudinal tilted fireball based on the Glauber collision geometry is presented.
This longitudinal tilted initial condition combined with the Ideal-CLVisc (3 + 1)D hydrodynamic model,
a nonvanishing directed flow coefficient $v_{1}$ in a wide rapidity range is observed.
After comparing the model's results with experimentally observed data of directed flow coefficient $v_{1}(\eta)$ from $\sqrt{s_{NN}}$ = 200 GeV Cu + Cu,
Au + Au collisions at RHIC energy to $\sqrt{s_{NN}}$ = 2.76 TeV and $\sqrt{s_{NN}}$ = 5.02 TeV Pb + Pb collisions at the LHC energy.
We find that the directed flow measurements in heavy-ion collisions can set strong constraints on the imbalance of forward and backward incoming nuclei
and on the magnitude asymmetry of pressure gradients along the $x$ direction.
\end{abstract}

\keywords{Relativistic Heavy-ion collisions, quark-gluon plasma, directed flow}
\pacs{25.75.Ld,25.75.Gz}

\maketitle
\date{\today}
\end{CJK*}
\section{Introduction}
\label{v1section1}
In the past decades, relativistic heavy-ion collisions provided methods to explore and understand the deconfined quark-gluon plasma (QGP).
The strongly collective flow of such hot dense medium is one of the key observations in the physics of high-energy heavy-ion collisions~\cite{Bass:1998vz,Gyulassy:2004zy,Shuryak:2003xe,Heinz:2013th}.

The rapidity-odd directed flow (shorted as directed flow or $v_{1}$) refers to the collective sideward deflection of particles
and is the first-order harmonic of the Fourier expansion of the particle azimuthal distribution with respect to the reaction plane~\cite{Ollitrault:1992bk,Voloshin:1994mz}.
The directed flow is believed to be created at very early stage (during the nuclear passage time: $2R/\gamma$~$\approx$ 0.1 fm/c) at large rapidity $\eta$ (in the fragmentation region)~\cite{Back:2004je}.
Therefore, it may keep trace of the bulk collective dynamics and the subsequent evolution into a thermalized hot QCD matter,
which provides a unique insight to investigate the initial condition.

At current stage, the directed flow $v_{1}$ and the splitting $\Delta v_{1}$ between particles
and anti-particles are measured for both light charged hadrons and heavy quark productions (\eg  $J/\psi$, $D^{0}$, $\bar{D}^{0}$)
at the BNL Relativistic Heavy Ion Collider (RHIC) and at the Large Hadron Collider (LHC)~\cite{Abelev:2008jga,Adamczyk:2014ipa,Adam:2019wnk,Acharya:2019ijj,Adamczyk:2017nxg}.
A striking characteristic of the measured directed flow for charged particle
is the large value of $v_{1}$ at RHIC energy~\cite{Abelev:2008jga} and
large value of splitting $\Delta v_{1}$ of charm hadrons produced at LHC energy~\cite{Acharya:2019ijj}.
The directed flow has been investigated by different models and mechanisms,
such as the transport model~\cite{Petersen:2006vm}, three dimensional (3D) initial geometric asymmetry
source + hydrodynamic model~\cite{Adil:2005qn,Bozek:2010bi,Chen:2019qzx,Shen:2020jwv,Ryu:2021lnx,Chatterjee:2017ahy,Chatterjee:2018lsx,Beraudo:2021ont,Heinz:2013th},
extremely strong magnetic-field effect~\cite{Gursoy:2014aka,Das:2016cwd,Gursoy:2018yai,Chatterjee:2018lsx,Sun:2020wkg,Dubla:2020bdz},
the chiral magnetic effect (CME) and magnetohydrodynamics (MHD)~\cite{Inghirami:2019mkc}, vorticity effect~\cite{Oliva:2020doe},
AMPT + quark coalescence model~\cite{Nasim:2018hyw}, and so on.

The CCNU-LBNL-viscous hydrodynamic (CLVisc) model~\cite{Pang:2018zzo} is an open source (3+1)-dimensional hydrodynamic frame
for heavy ion collisions that was developed by {\bf \href{https://inspirehep.net/authors/1274264}{{\tt \emph{L-G Pang}}}} et al.,
which is parallelized on a graphics processing unit (GPU) using the Open Computing Language (OpenCL).
There is tremendous progress in understanding the QCD matter by using the CLVisc model, such as the strong vorticity prediction~\cite{Pang:2016igs},
which was later found at the RHIC-STAR~\cite{STAR:2017ckg}, the magnetic-field-induced squeezing effect calculation~\cite{Pang:2016yuh},
the description for longitudinal decorrelation of anisotropic flow~\cite{Pang:2015zrq,Wu:2018cpc},
deep learning, and machine learning coupled with heavy ion collisions~\cite{Pang:2016vdc}, and the jet quenching research~\cite{Chen:2017zte,He:2018gks}.
The original configurations and code of CLVisc (3+1)D hydrodynamic can be downloaded available in gitlab website:
{\bf [\href{https://gitlab.com/snowhitiger/PyVisc}{{\tt https://gitlab.com/snowhitiger/PyVisc}}]}

In this paper, {\color{black} following the well-known Boz$\dot{\textrm{e}}$k-Wyskiel parametrization tilted initial condition in Refs.~\cite{Bozek:2011ua,Bozek:2010bi,Inghirami:2019mkc,Beraudo:2021ont},
an alternative parametrization to generate the tilted longitudinal structure of the fireball is presented.}
The contribution of the forward-going and backward-going participant nucleons is assumed to be imbalance and related to a phenomenological parameter $H_{t}$ in the Glauber model.
Such a modified initial condition with the Ideal-CLVisc (3 + 1)D hydrodynamics simulation give a finite directed flow of the charged particles (and $\pi^{+}$)
in the middle and backward/forward rapidity region compared with the STAR and ALICE measurements.

The article is organized as follows. In Sec.~\ref{v1section2}, the parametrization and the modified initial condition
for the CLVisc(3+1)D ideal hydrodynamic simulation are presented.
In Sec.~\ref{v1section3}, our numerical results on the directed flow of charged particles are presented.
Finally, in Sec.~\ref{v1section4}, we summarize the results and present a short outlook.

\section{Setup of the numerical simulations}
\label{v1section2}

{\color{black}
In this work, following the Refs.~\cite{Bozek:2010bi,Inghirami:2019mkc,Beraudo:2021ont},
the viscous corrections are not included in the present study
\footnote{{\color{black}
The directed flow is generated at very early stage in the evolution and it is given by the initial tilted source.
The ideal fluid approximation could give a sizable directed flow coefficient $v_{1}$ and directly reflect the evolution of pressure~\cite{Bozek:2010bi}.}}
and the viscosity effect will be included in future studies. The (3 + 1)-dimensional numerical simulations are performed with in Milne/Bjorken coordinates.
}

\subsection{Initial condition}

\label{ICond}
The initial energy density distribution is computed according to a modified optical Glauber model~\cite{Miller:2007ri,Pang:2018zzo}.

In the optical Glauber model, the nucleus thickness function $T(x,y)$ from the Woods-Saxon distribution is
\begin{equation}
\begin{aligned}
T(x,y)=\int_{-\infty}^{\infty}dz\frac{n_{0}}{1+e^{(\sqrt{x^{2}+y^{2}+z^{2}}-R)/d}}
\label{lorentz:Fx}
\end{aligned}
\end{equation}
where $n_{0}$ is the average nuclear density, $d$ is the diffusiveness, $x,~y,~z$ is the space coordinates and $R$ is the radius of the nuclear
Fermi distribution, which depends on the specific nucleus. The parameters used for nucleus Cu, Au and Pb in current study are listed in Table.~\ref{t:parameters}.
\begin{table}[!h]
\begin{center}
\begin{tabular}{l c c c c}
\hline\hline
Nucleus     & A          & $n_{0}$ [1/fm$^{3}$]      & R~[fm]    & $d$ [fm]    \\ \hline
Cu          & 63         & 0.17                      & 4.20      & 0.546       \\
Au          & 197        & 0.17                      & 6.38      & 0.546       \\
Pb          & 208        & 0.17                      & 6.62      & 0.535       \\
\hline\hline
\end{tabular}
\caption{\label{t:parameters} Table of parameters used in the Woods-Saxon distribution for Cu, Au and Pb described in the text~\cite{Loizides:2017ack}.}
\end{center}
\end{table}

$T_{1}(\mathbf{x}_{T})$ and $T_{2}(\mathbf{x}_{T})$ are the densities of participants from the two nuclei,
\begin{equation}
\begin{aligned}
T_{1}(\mathbf{x}_{T})=T_{+}(\mathbf{x}_{T})\left(1-\left(1-\frac{\sigma_{NN} T_{-}(\mathbf{x}_{T})}{A}\right)^{A}\right)
\label{eq:t1}
\end{aligned}
\end{equation}
\begin{equation}
\begin{aligned}
T_{2}(\mathbf{x}_{T})=T_{-}(\mathbf{x}_{T})\left(1-\left(1-\frac{\sigma_{NN} T_{+}(\mathbf{x}_{T})}{A}\right)^{A}\right)
\label{eq:t2}
\end{aligned}
\end{equation}
where $A$ is the mass number of the colliding nuclei, $\sigma_{NN}$ is the inelastic-scattering cross section, $\sigma_{NN}$ are set to 40 mb for $\sqrt{s_{NN}}$ = 200 GeV Cu + Cu,
Au + Au, 64mb for $\sqrt{s_{NN}}$ = 2.76 TeV Pb + Pb and 67mb for Pb + Pb $\sqrt{s_{NN}}$ = 5.02 TeV collisions~\cite{Loizides:2017ack}, and
\begin{equation}
\begin{aligned}
T_{+}(\mathbf{x}_{T})=T(\mathbf{x}_{T}+\mathbf{b}/2),~~~~T_{-}(\mathbf{x}_{T})=T(\mathbf{x}_{T}-\mathbf{b}/2)
\label{eq:t+}
\end{aligned}
\end{equation}
where $\mathbf{x}_{T}~=~(x,y)$ is the vector of the transverse plane coordinates and $\mathbf{b}$ is the impact-parameter vector, connecting
the centers of the two nuclei.

The function $W_{N}$ gives the contribution of the wounded nucleons.
To generate a tilted fireball along the longitudinal direction,
the wounded nucleons~\cite{Miller:2007ri} weight function $W_{N}$ is modified as follow,
\begin{equation}
\begin{aligned}
W_{N}(x,y,\eta_{s})=&[T_{1}(x,y)+T_{2}(x,y)] \\
&+H_{t}[T_{2}(x,y)-T_{1}(x,y)]\tan\left(\frac{\eta_{s}}{\eta_{t}}\right),
\label{eq:mwneta1}
\end{aligned}
\end{equation}
here the longitudinal tilted parameter $H_{t}$ is a free parameter that reflects the imbalance in the emitting contributions from forward-going and backward-going participant nucleons.
As we will see in the following section, varying the parameter $H_{t}$ results in strong 
dependencies in the magnitude of pion's and charged particle's directed flow.
The parameter $\eta_{t}$ is a constant and chosen to be 8.0 for all the collision systems.

The energy density distribution at the hydrodynamic starting time $\tau_{0}$ is given by
\begin{equation}
\begin{aligned}
\varepsilon(x,y,\eta_{s})=\varepsilon_{0} \cdot W(x,y,\eta_{s})H(\eta_{s}),
\label{eq:ep1}
\end{aligned}
\end{equation}
where $\varepsilon_{0}$ the maximum energy density given in Table.~\ref{t:epslion0}.
For most-central collisions at RHIC and the LHC energy, the total weight function $W(x,y,\eta_{s})$ is defined as~\cite{Pang:2016yuh,Inghirami:2019mkc}
\begin{equation}
\begin{aligned}
W(x,y,\eta_{s})=\frac{(1-\alpha)W_{N}(x,y,\eta_{s})+\alpha n_{BC}(x,y)}{(1-\alpha)W_{N}(0,0,0)+\alpha n_{BC}(0,0)|_{\mathbf{b}=0}},
\label{eq:wneta}
\end{aligned}
\end{equation}
with $\alpha=0.05$ being the collision hardness parameter\footnote{The parameter $\alpha$ should be collision energy dependent
when reproducing the centrality dependence of multiplicity necessitates. In this paper we follow Refs.~\cite{Inghirami:2019mkc,Pang:2016yuh} and assume $\alpha=0.05$ for simplicity. } and $n_{BC}(x,y)$ being the mean number of binary collision:
\begin{equation}
\begin{aligned}
n_{BC}(x,y)=\sigma_{NN}T_{+}(x,y)T_{-}(x,y).
\label{eq:nbc}
\end{aligned}
\end{equation}

The energy density profile in the longitudinal direction is modulated by~\cite{Pang:2018zzo},
\begin{equation}
\begin{aligned}
H(\eta_{s})=\exp\left[-\frac{(\eta_{s}-\eta_{w})^{2}}{2\sigma^{2}_{\eta}}\theta(\eta_{s}-\eta_{w}) \right],
\label{eq:eta}
\end{aligned}
\end{equation}
while the parameters $\eta_{w}$ and $\sigma_{\eta}$ are set to
2.95 and 0.4 for Au + Au and Cu + Cu at $\sqrt{s_{NN}}=200$ GeV, 3.6 and 0.6 for Pb + Pb at $\sqrt{s_{NN}}=2.76$ TeV and 3.6 and 0.7 for Pb + Pb at $\sqrt{s_{NN}}=5.02$ TeV
collisions. $\eta_{w}$ and $\sigma_{\eta}$ lead to a two-peak structure along the longitudinal direction in the final state~\cite{Pang:2016yuh,Pang:2018zzo,Inghirami:2019mkc}.

\textrm{}
\begin{table}[!h]
\begin{center}
\begin{tabular}{l c c c c}
\hline\hline
System      & $\sqrt{s_{NN}}$   & $\varepsilon_{0}$ [GeV/fm$^{3}$]\\ \hline
Cu+Cu       & 200 GeV                    & 83.5                        \\
Au+Au       & 200 GeV                    & 155.5                       \\
Pb+Pb       & 2.76 TeV                    & 465.0                      \\
Pb+Pb       & 5.02 TeV                    & 580.0                       \\
\hline\hline
\end{tabular}
\caption{\label{t:epslion0} Maximum energy density for CLVisc (3 + 1)D ideal hydrodynamics starting from initial proper time $\tau_{0}=0.2$ fm
to reproduce the charged multiplicity distribution at RHIC and at the LHC~\cite{Pang:2016yuh}.}
\end{center}
\end{table}
The collision centrality class is determined by the impact parameter $b$, 
which can be obtained by interpolation~\cite{Loizides:2017ack}.
The impact parameter using in current study is presented in Table.~\ref{t:impactp}.
\begin{table}[!h]
\begin{center}
\begin{tabular}{l c c c c c c c}
\hline\hline
~~~ $b$    & 0-5\%     & 10-15\%    & 10-20\%          & 30-40\%  & 30-60\%   & 5-40\% \\ \hline
Cu+Cu      & 1.74      & 3.70       & 4.03             & 6.17     & 6.85      & 4.78\\
Au+Au      & 2.40      & 5.27       & 5.76             & 8.78     & 9.76      & 6.73\\
Pb+Pb      & 2.65      & 5.58       & 6.09             & 9.33     & 10.28     & 7.18\\
\hline\hline
\end{tabular}
\caption{\label{t:impactp} Impact parameters $b$ used in the Glauber model for Cu + Cu, Au + Au and Pb + Pb described in the text~\cite{Loizides:2017ack}.}
\end{center}
\end{table}

\begin{figure}[!ht]
\begin{center}
\includegraphics[width=0.85\linewidth]{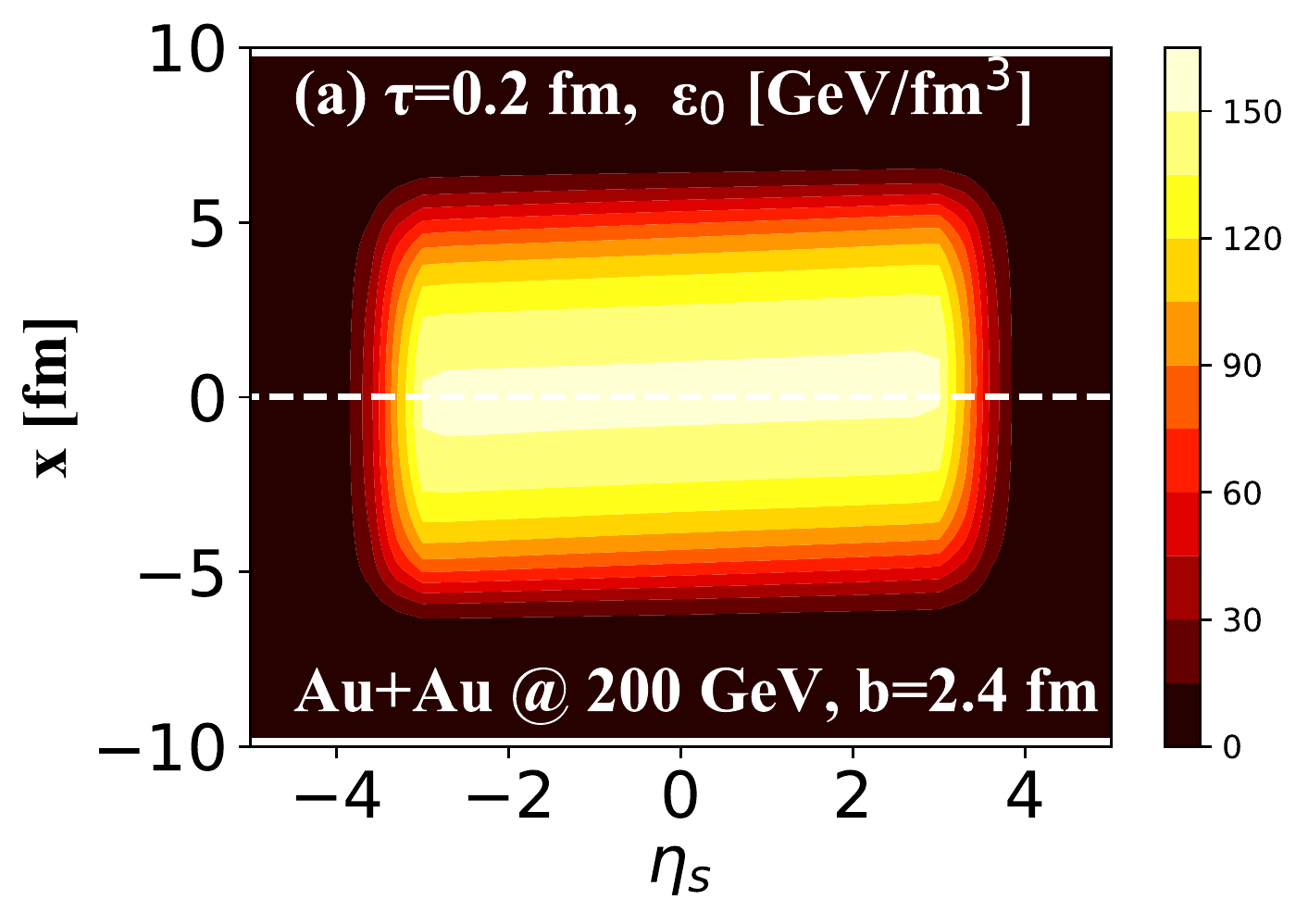}~\\
\includegraphics[width=0.85\linewidth]{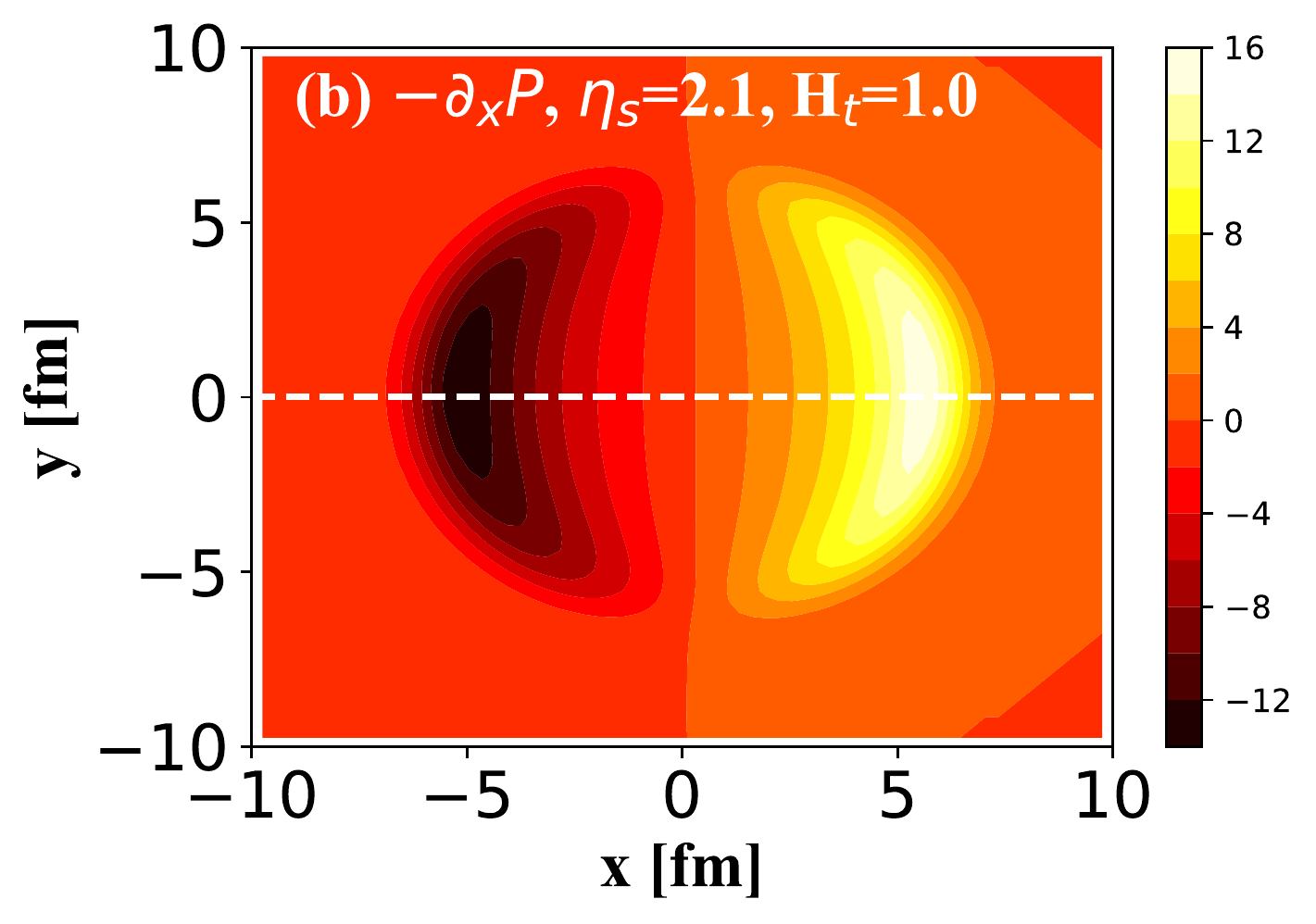}~\\
\includegraphics[width=0.85\linewidth]{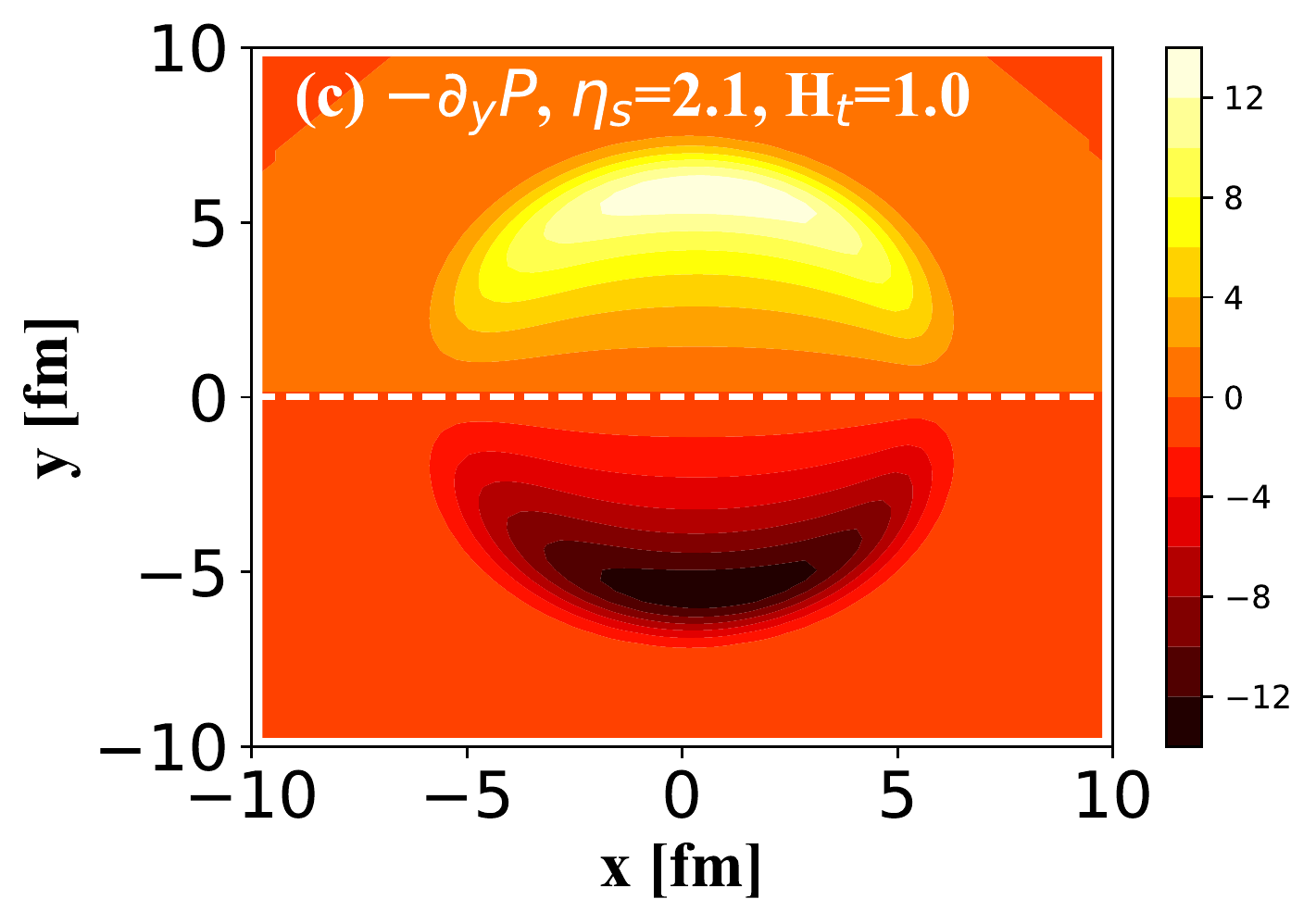}~
\end{center}
\caption{(Color online) Color contours show the initial condition at $\tau$ = 0.2 fm in 0-5\% Au+Au collisions at $\sqrt{s_{NN}}~=~200$ GeV.
(top panel) The initial energy density profile on the $\eta_{s}-x$  plane.
(middle and bottom panel) The magnitude of pressure gradients $-\partial_{x}P$ and $-\partial_{y}P$ at $\tau_{0}=0.2$ fm and $\eta_{s}$ = 2.1.
The impact parameter $b$ = 2.4 fm is consistent with the centrality class 0-5\%.}
\label{f:auau200pressure}
\end{figure}

\begin{figure}[!ht]
\begin{center}
\includegraphics[width=0.85\linewidth]{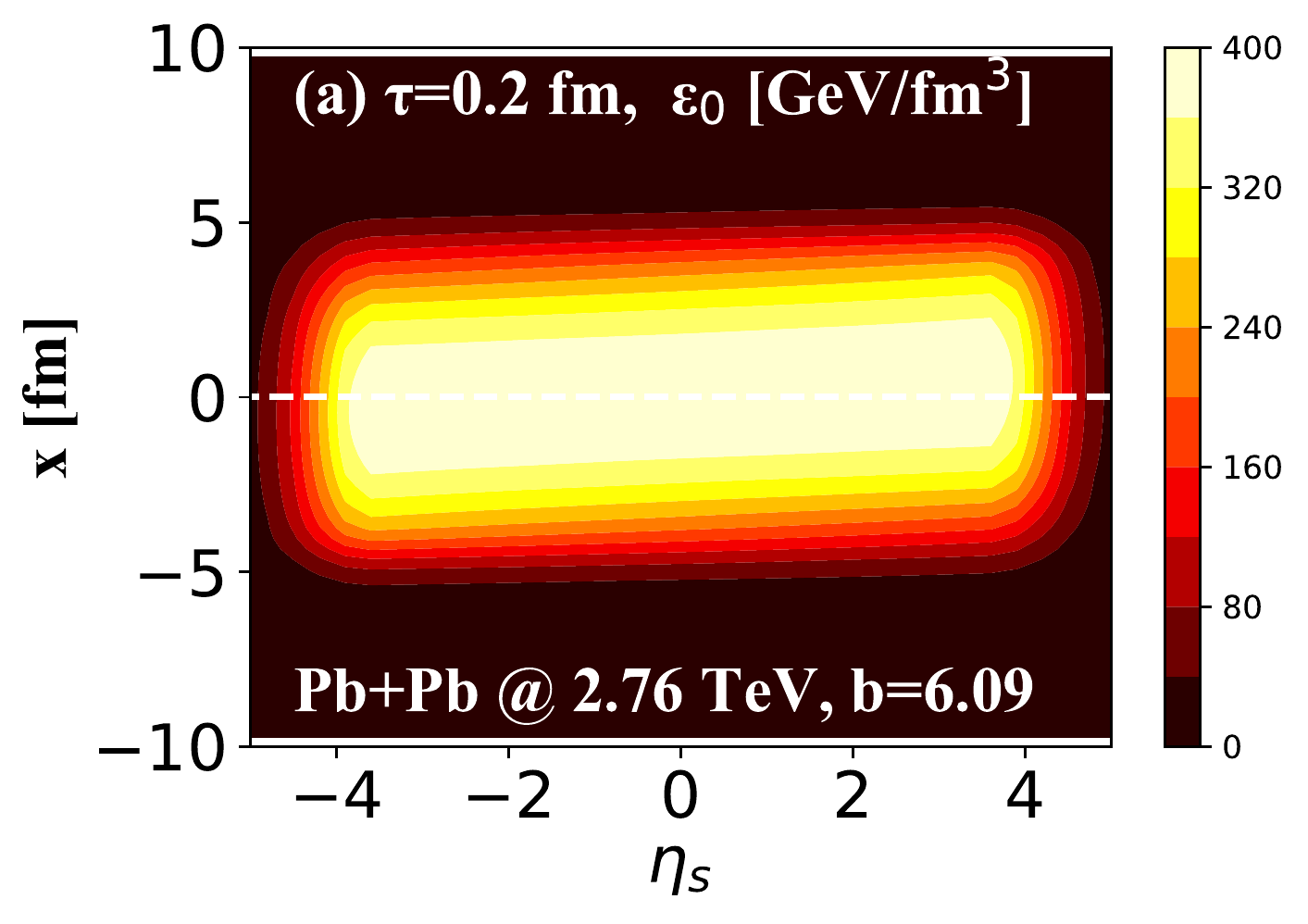}~\\
\includegraphics[width=0.85\linewidth]{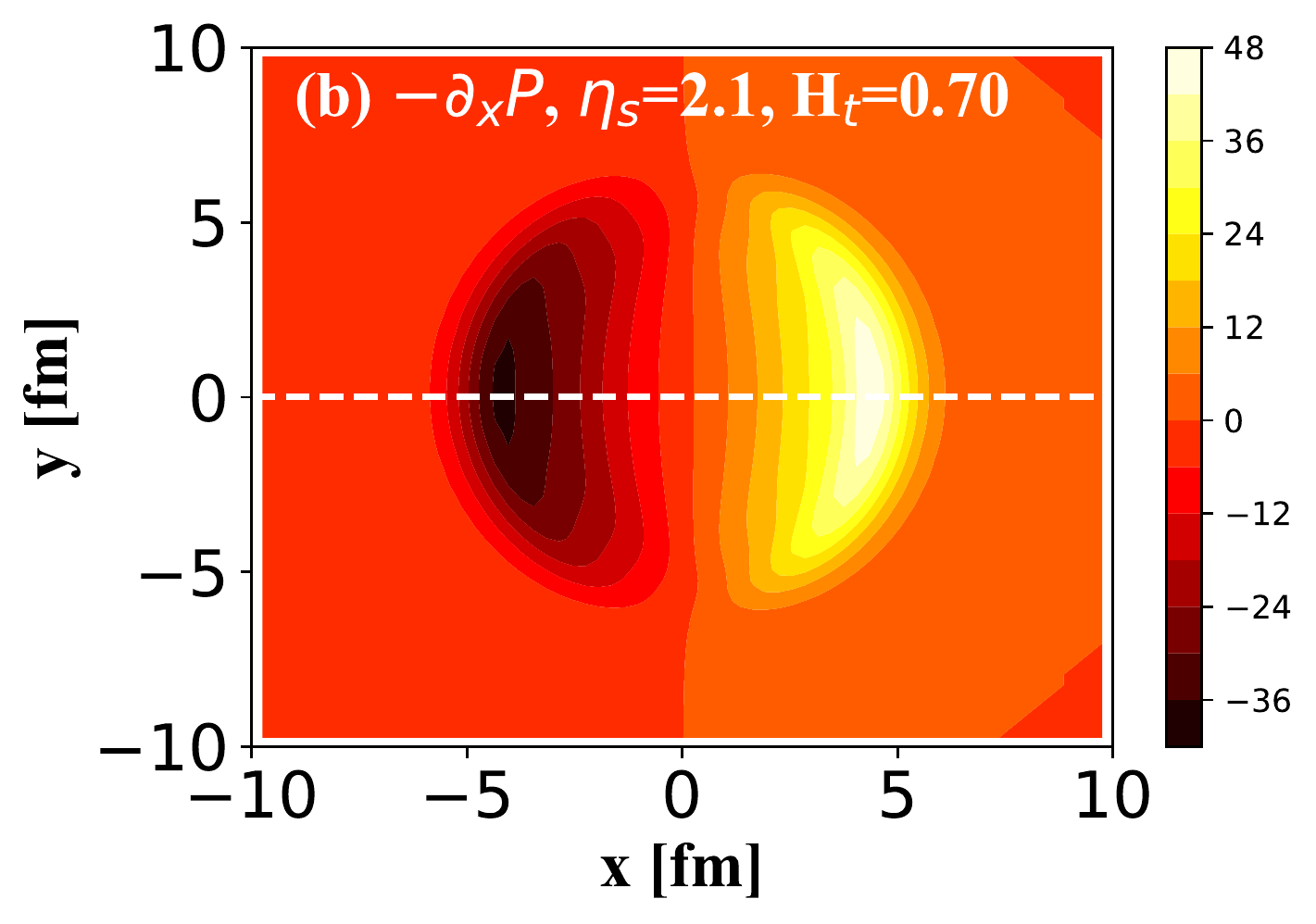}~\\
\includegraphics[width=0.85\linewidth]{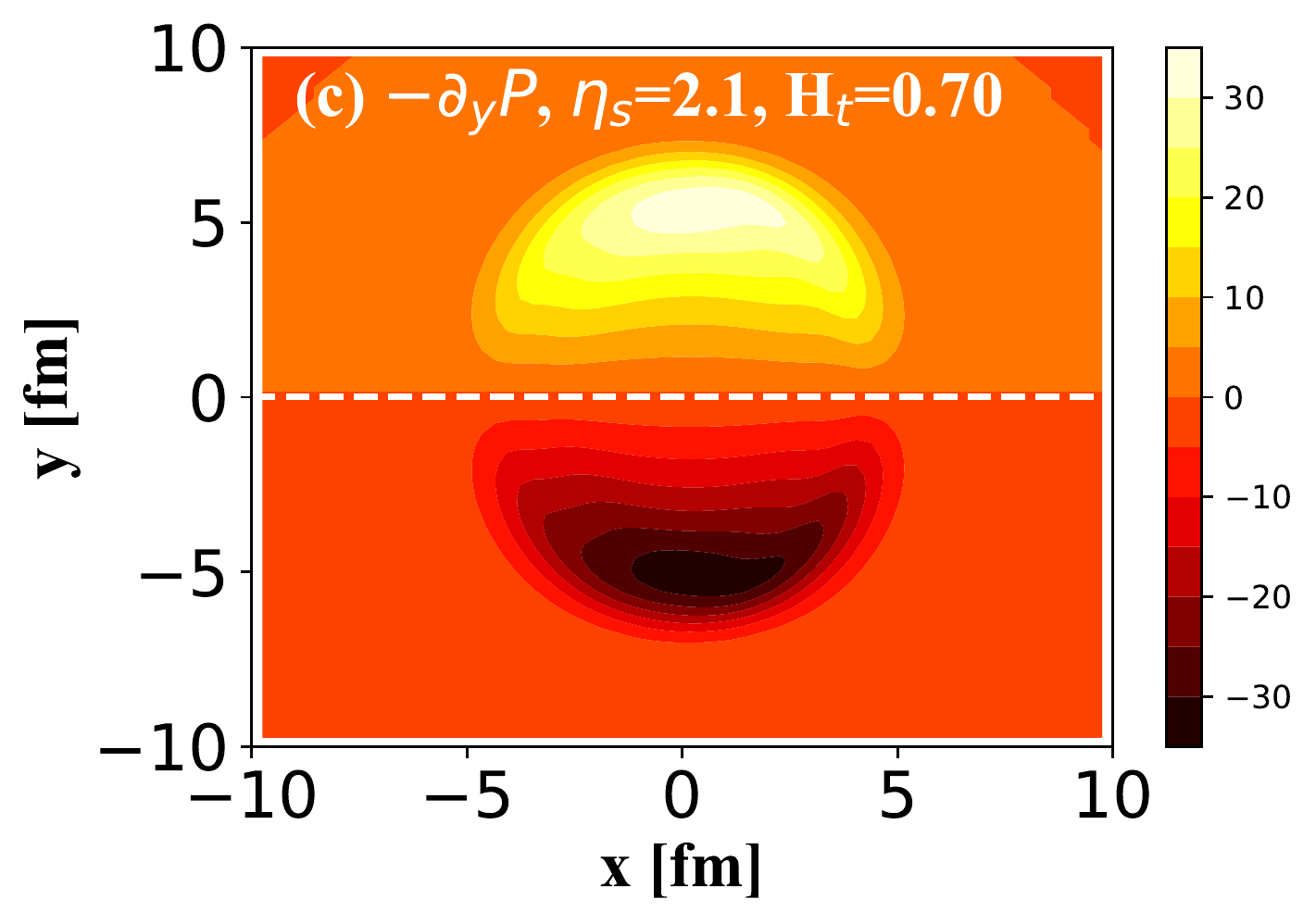}~
\end{center}
\caption{Color contour plot for initial condition at $\tau$ = 0.2 fm in 10\%-20\% Pb+Pb collisions at $\sqrt{s_{NN}}=2.76$ TeV.
(top panel) The initial energy density profile on the $\eta_{s}-x$ plane (at y = 0.0 fm).
(middle and bottom panel) The  magnitude of pressure gradients with H$_{t}$ = 0.70 at $\tau_{0}=0.2$ fm and $\eta_{s}$ = 2.1.
The impact parameter $b$ = 6.09 fm is consistent with the centrality class 10\%-20\%.}
\label{f:pbpbpressure}
\end{figure}

With the above parametrizations, we illustrate the profile of energy density and magnitude of pressured gradient distribution 
for  0-5\% Au + Au collisions at $\sqrt{s_{NN}}$ =  200 GeV
and 10-20\% Pb + Pb collisions at $\sqrt{s_{NN}}$ = 2.76 TeV in Figs.~\ref{f:auau200pressure} and~\ref{f:pbpbpressure}.

Figure~\ref{f:auau200pressure}(a) shows the energy density distribution on the $\eta_{s}-x$ plane
(y = 0.0 fm) at $\tau$ = 0.2 fm in centrality class 0-5\% ($b$ = 2.4 fm) Au + Au collisions
at $\sqrt{s_{NN}}$ =  200 GeV with the tilted parameter $H_{t}$ = 1.0.
It is evident that the parameter H$_{t}$ in Eq.~(\ref{eq:mwneta1}) controls the imbalance
in the forward-backward hemispheres and leads to a longitudinal tilted fireball.

Figures~\ref{f:auau200pressure}(b) and \ref{f:auau200pressure}(c) show the magnitude of initial pressure gradients in the transverse plane at $\tau$ = 0.2 fm
and forward-rapidity $\eta_{s}=2.1$.
The magnitude of pressure gradient ($-\partial_{x}P$ and $-\partial_{y}P$) is calculated from the initial energy density and equation of state (EoS). We find that
the magnitude of the pressure gradient shows an asymmetry along the $x>0$ and $x<0$ direction in the transverse plane.
The value of $-\partial_{x}P$ shows a maximal value of around 16 GeV/fm$^{4}$ (14 GeV/fm$^{4}$) at the $x>0$ ($x<0$) panel.
With the hydrodynamic expansion, considering the contribution from the transverse and longitudinal pressures,
the final directed flow coefficient becomes negative for positively rapidity~\cite{Bozek:2011ua,Bozek:2010bi}.

Figure~\ref{f:pbpbpressure}(a)
shows the initial energy density distribution on the $\eta_{s}-x$ plane (y = 0.0 fm) at $\tau$ = 0.2 fm in 10\%-20\% ($b$ = 6.09) Pb + Pb collisions
at $\sqrt{s_{NN}}$ = 2.76 TeV with the parameter $H_{t}$ = 0.70.

Figure~\ref{f:pbpbpressure}(b) and \ref{f:pbpbpressure}(c) show the magnitude of initial pressure gradients in the transverse plane
at $\tau$ = 0.2 fm and forward-rapidity $\eta_{s}=2.1$.
The magnitude of pressure gradient ($-\partial_{x}P$ and $-\partial_{y}P$) shows an asymmetry along the $x>0$ and $x<0$ direction. Furthermore, the value of $-\partial_{x}P$
shows a maximal value of around 48 GeV/fm$^{4}$ (40 GeV/fm$^{4}$) along the $x>0$ ($x<0$) direction in the transverse plane.

The information on how the above initial spatial anisotropy is transferred to the momentum space~\cite{Bozek:2011ua,Bozek:2010bi} is encoded on the directed flow coefficient,
which will be presented in Sec.~\ref{v1section3}.

\subsection{Hydrodynamic equations/simulation}
\label{hydro}
The hydrodynamic equations from the literature read~\cite{Jiang:2020big,Jiang:2018qxd,Denicol:2012cn,Romatschke:2009im,Romatschke:2017ejr}
\begin{equation}
\begin{aligned}
\partial_{\mu}T^{\mu\nu}=0,
\label{eq:tmn}
\end{aligned}
\end{equation}
where $T^{\mu\nu}=\varepsilon u^{\mu}u^{\nu}-P\Delta^{\mu\nu}$ is the energy-momentum tensor for ideal hydrodynamics, $u^{\mu}=\gamma$(1, \textbf{v}) denotes
the fluid velocity four-vector, $\varepsilon$ is the energy density and $P$ is the pressure. The pressure $P$ is given as a function of energy density $\varepsilon$
by the equation of state (EoS). The lattice QCD equation of state from the Wuppertal-Budapest group (2014)~\cite{Borsanyi:2013bia} is used in the current study.
Notice that the viscosity and net baryon density are set to zero in current study.

The energy-momentum conservation equations~[Eq. (\ref{eq:tmn})] are solved numerically
by using Kurganov-Tadmor (KT) algorithm~\cite{Pang:2018zzo}, which was introduced
to the field of high-energy physics by the McGill group~\cite{Schenke:2010nt}.
The 3D partial differential equations are solved by updating the values of fluid cells at each time step.
For each collision system, we run the Ideal-CLVisc (3 + 1)D hydrodynamic simulation with number of
cells $N_{\rm{cells}}=N_{x}\times N_{y} \times N_{\eta_{s}}=201 \times 201 \times 105$ for 1600 time steps on GPU NVIDIA GeForce RTX 2080TI (Turing Features)
and server CPU Intel Xeon E5-1620v2. For more details about this GPU simulation part, see Refs.~\cite{Pang:2018zzo,Du:2019civ}.
The freeze-out condition used in current calculation is the isothermal freeze-out condition~\cite{Pang:2018zzo},
where one assumes the hypersurface is determined by a constant temperature $T_{frz}=137$ MeV.

Based on the Cooper-Frye formula~\cite{Cooper:1974mv}, the Ideal-CLVisc provided a ``smooth method'' to compute the different particle spectra on the freeze-out hypersurface,
where the numerical integration is performed over the freeze-out hyper-surface and smooth particle spectra are obtained
in $N_{Y}~\times~N_{pt}~\times~N_{\phi}$ = 41$\times$ 15 $\times$ 48 tabulated ($Y,~p_{T},\phi$) bins~\cite{Pang:2018zzo}.
$p_{T}$ and $\phi$ are chosen to be Gaussian quadrature nodes to simplify the calculation of $p_{T}$ or $\phi$ integrated spectra.
Hadron spectra from resonance decays are also computed via integration and parallelized on the GPU.

\subsection{Rapidity-odd directed flow coefficient $v_{1}(\eta)$}
\label{observables}
Directed flow coefficient $v_{1}(\eta)$ reflects the collective sideward deflection of particles.
Here $v_{1}(\eta)$ is calculated via integration as follows,
\begin{equation}
\begin{aligned}
v_{1}(\eta)=\langle\cos(\phi-\Psi_{1})\rangle=\frac{\int\cos(\phi-\Psi_{1})\frac{dN}{d\eta d\phi}d\phi}{\int\frac{dN}{d\eta d\phi}d\phi},
\label{eq:v1}
\end{aligned}
\end{equation}
where $\Psi_{1}$ is the first-order event plane of the collision~\cite{Bozek:2010bi}.

\section{Numerical Results}
\label{v1section3}
In this section, numerical results from ideal CLVisc hydrodynamic simulation with experimental data for RHIC and the LHC energies are presented.

\subsection{Cu+Cu and Au+Au $\sqrt{s_{NN}}$ = 200 GeV collisions}

\begin{figure}[!htb]
\begin{center}
\includegraphics[width=0.8\linewidth]{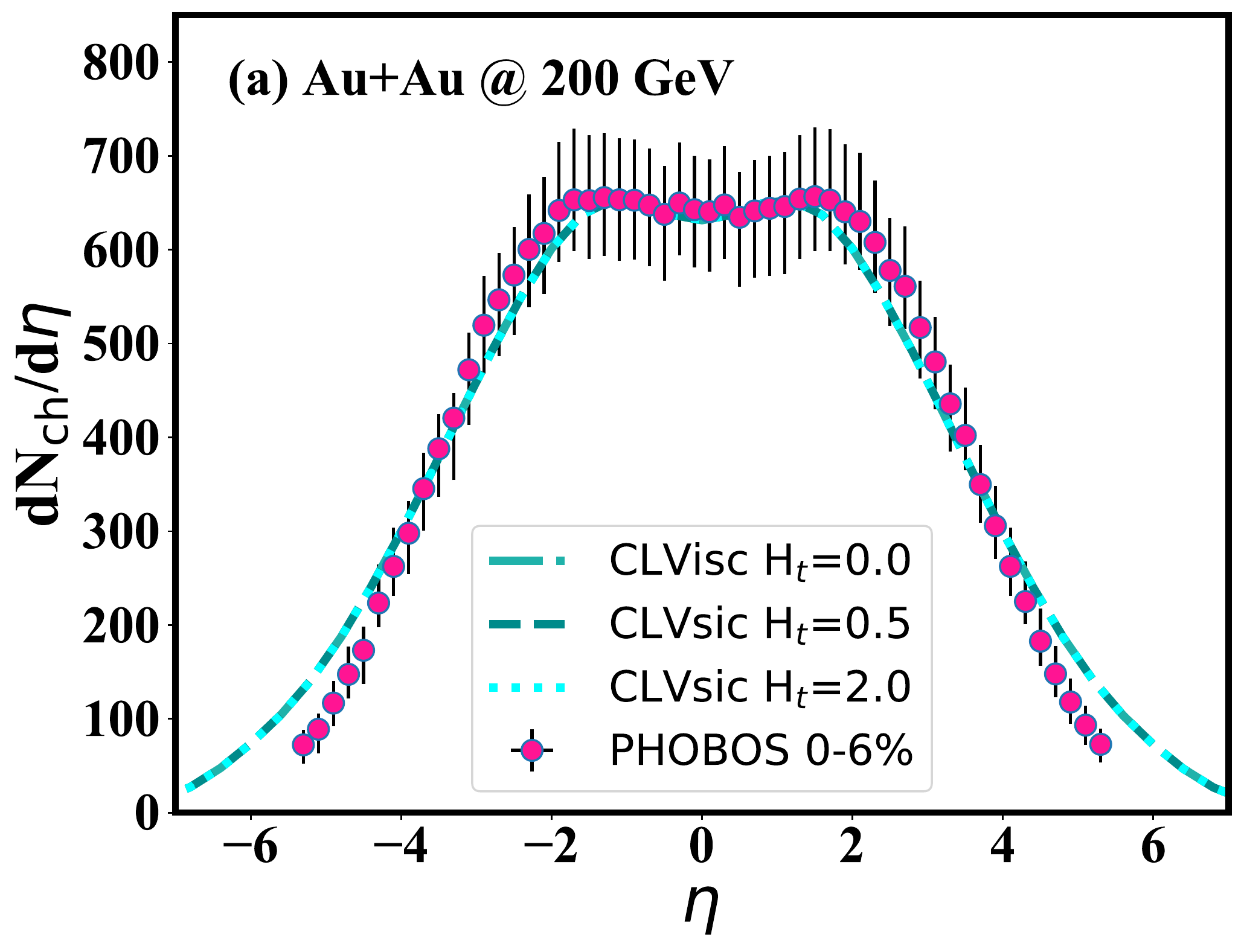}~\\
\includegraphics[width=0.8\linewidth]{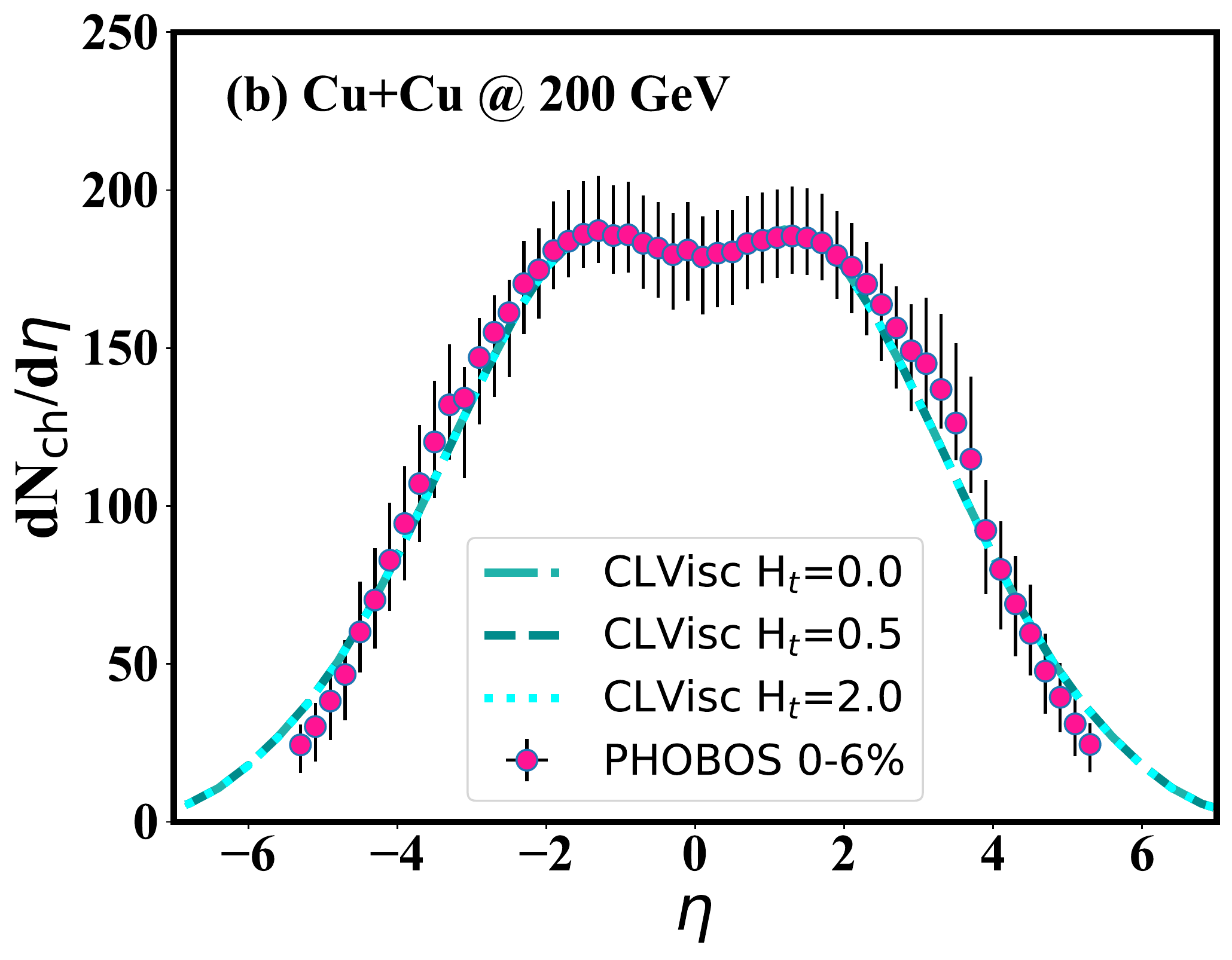}~
\end{center}
\caption{(Color online) The charged particle pseudorapidity distribution from CLVsic for
(a) Au + Au $\sqrt{s_{NN}}=200$ GeV collisions and (b) Cu + Cu $\sqrt{s_{NN}}=200$ GeV collisions in comparison to the experiment data
from PHOBOS Collaboration at top RHIC energy~\cite{Alver:2010ck} with three different settings of H$_{t}$ in the centrality class 0-6\%.}
\label{f:rhicdn}
\end{figure}

Figure~\ref{f:rhicdn} shows a comparison of charged hadron pseudo rapidity distributions $dN/d\eta$ between our model
and the PHOBOS measurement~\cite{Alver:2010ck} in the most central centrality 0-6\%\footnote{If one modifies the impact parameter $b$
and makes it consistent with the specified centrality class, then the model can fitting the multiplicity distribution dN$_{ch}$/d$\eta$ well.} Au+Au and Cu+Cu collisions at $\sqrt{s_{NN}}=200$ GeV.
We find that the dN/d$\eta$ obtained from the Ideal-CLVisc with modified initial conditions under three different settings of $H_{t}$ are almost indistinguishable on the plot.

\begin{figure}[!htb]
\begin{center}
\includegraphics[width=0.85\linewidth]{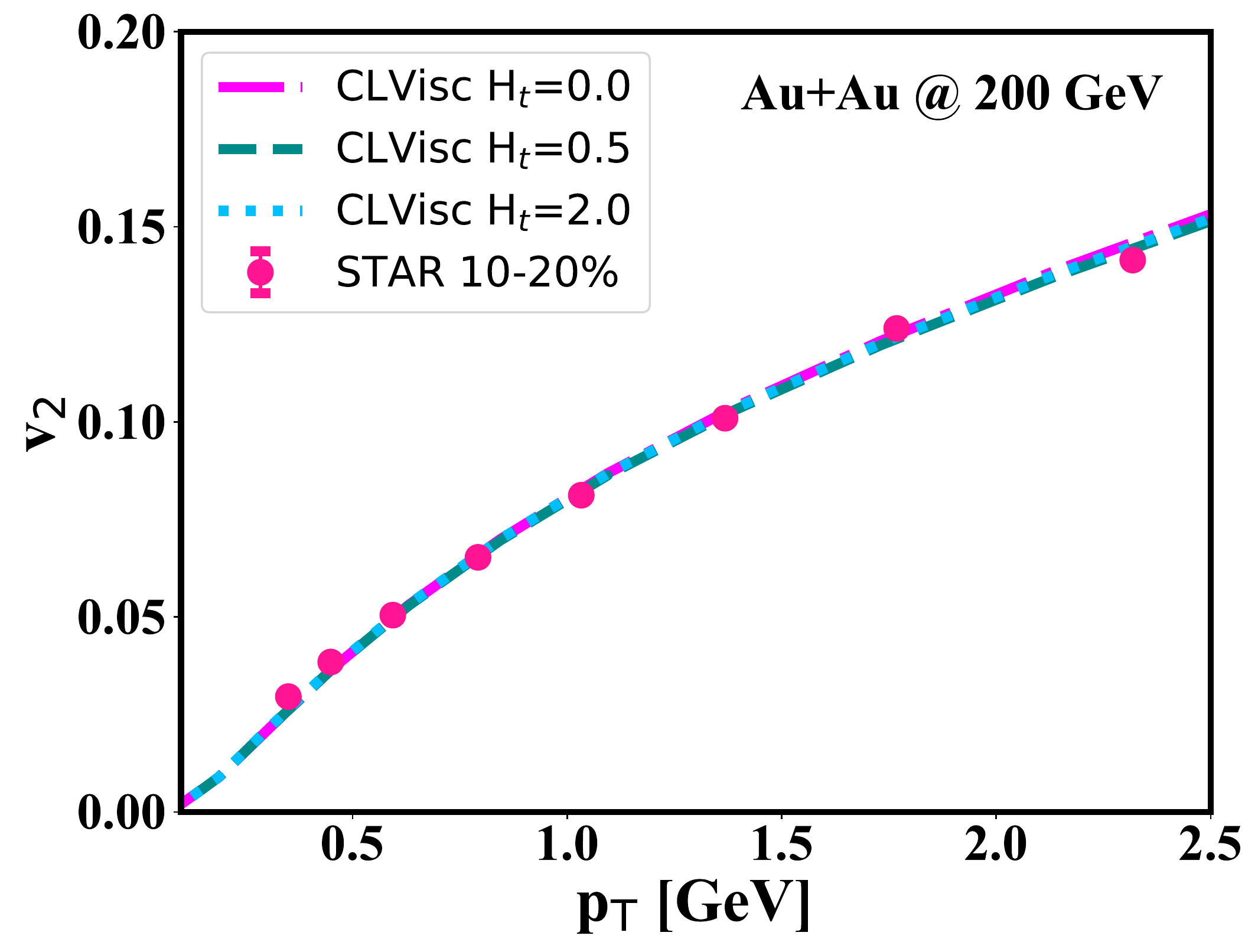}\\
\includegraphics[width=0.85\linewidth]{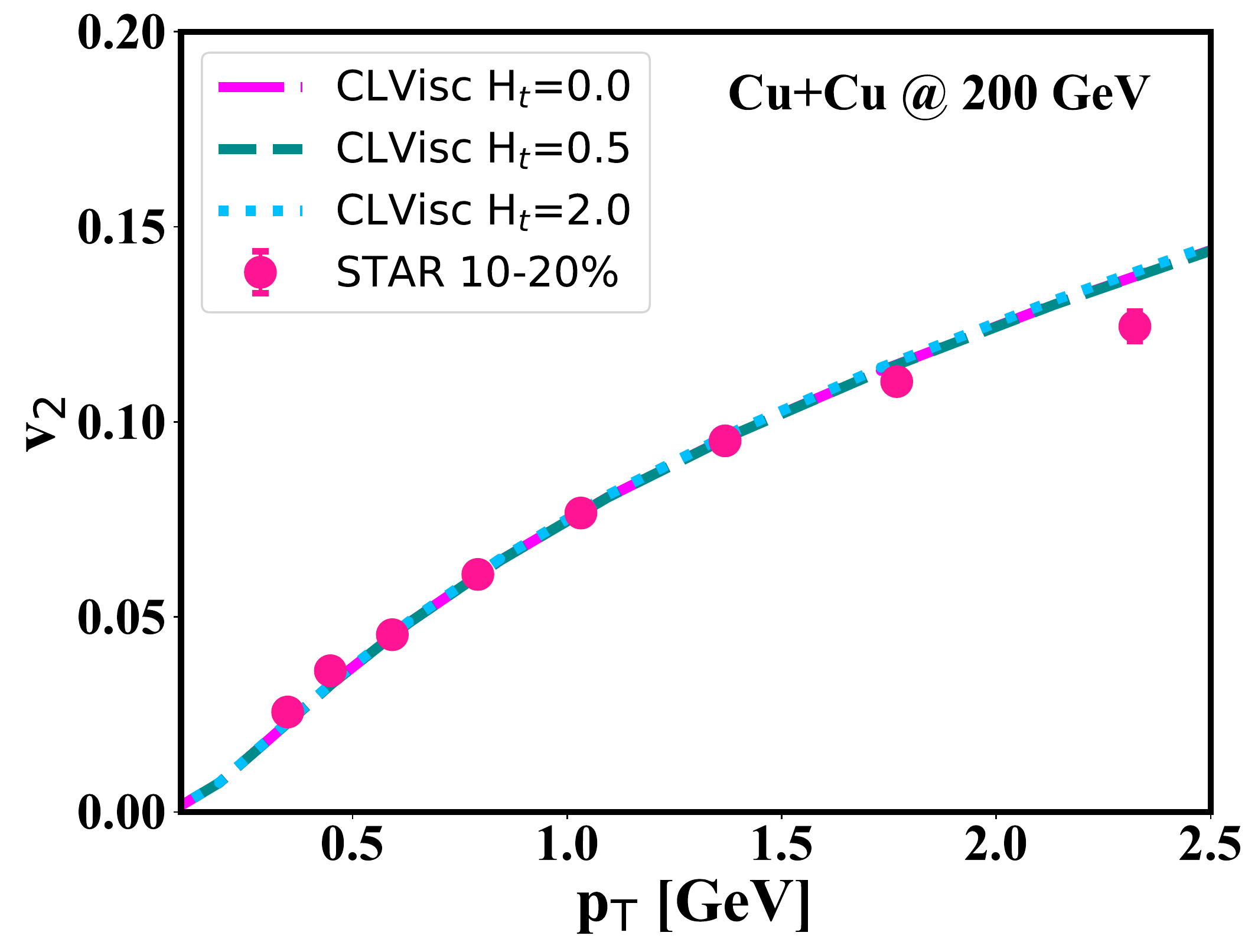}~
\end{center}
\caption{(Color online) Elliptic flow coefficients $v_{2}(p_{\textrm{T}})$ of charged particles at RHIC energy for the centrality class 10\%-20\%.
 (top panel) Au+Au $\sqrt{s_{NN}}=200$ GeV collisions. (bottom panel) Cu+Cu $\sqrt{s_{NN}}=200$ GeV collisions.
 The experimental data come from the STAR Collaboration~\cite{Adams:2004bi}.}
\label{f:richv2}
\end{figure}

Figure~\ref{f:richv2} shows elliptic flow coefficients $v_{2}(p_{\textrm{T}})$ of the charged particle in only 10\%-20\%
 Au+Au (top panel) and Cu+Cu (bottom panel) collisions at $\sqrt{s_{NN}}$ = 200 GeV
\footnote{Please notice that our model following those assumptions:
(1) modified optical Glauber model, which miss event-by-event eccentricity fluctuations; (2) dissipative effect is not included during the hydro expansion;
(3) the freeze-out temperature $T_{frz}$ = 137 MeV. Above assumptions makes our model sightly overestimate the elliptic
flow coefficient $v_{2}(p_{\textrm{T}})$ and $v_{2}(\eta)$ in several centrality classes at current stage~\cite{Shen:2020jwv,Hirano:2002ds,Hirano:2005xf}.
We will focus on the event-by-event dissipative hydrodynamic simulations
in future work.}.
Our hydrodynamic simulations with modified initial conditions give a same distribution of $v_{2}(p_{\textrm{T}})$ under different $H_{t}$ compared with the STAR measurement~\cite{Adams:2004bi}.

Our theoretical calculations of pseudorapidity distribution $dN/d\eta$ and elliptic flow $v_{2}(p_{\textrm{T}})$ for different settings of tilted parameter $H_{t}$ show
that different initial longitudinal tilted fireball almost does not affect the pseudo rapidity distribution and the elliptic flow coefficient $v_{2}(p_{T})$.

\begin{figure}[!htb]
\begin{center}
\includegraphics[width=0.85\linewidth]{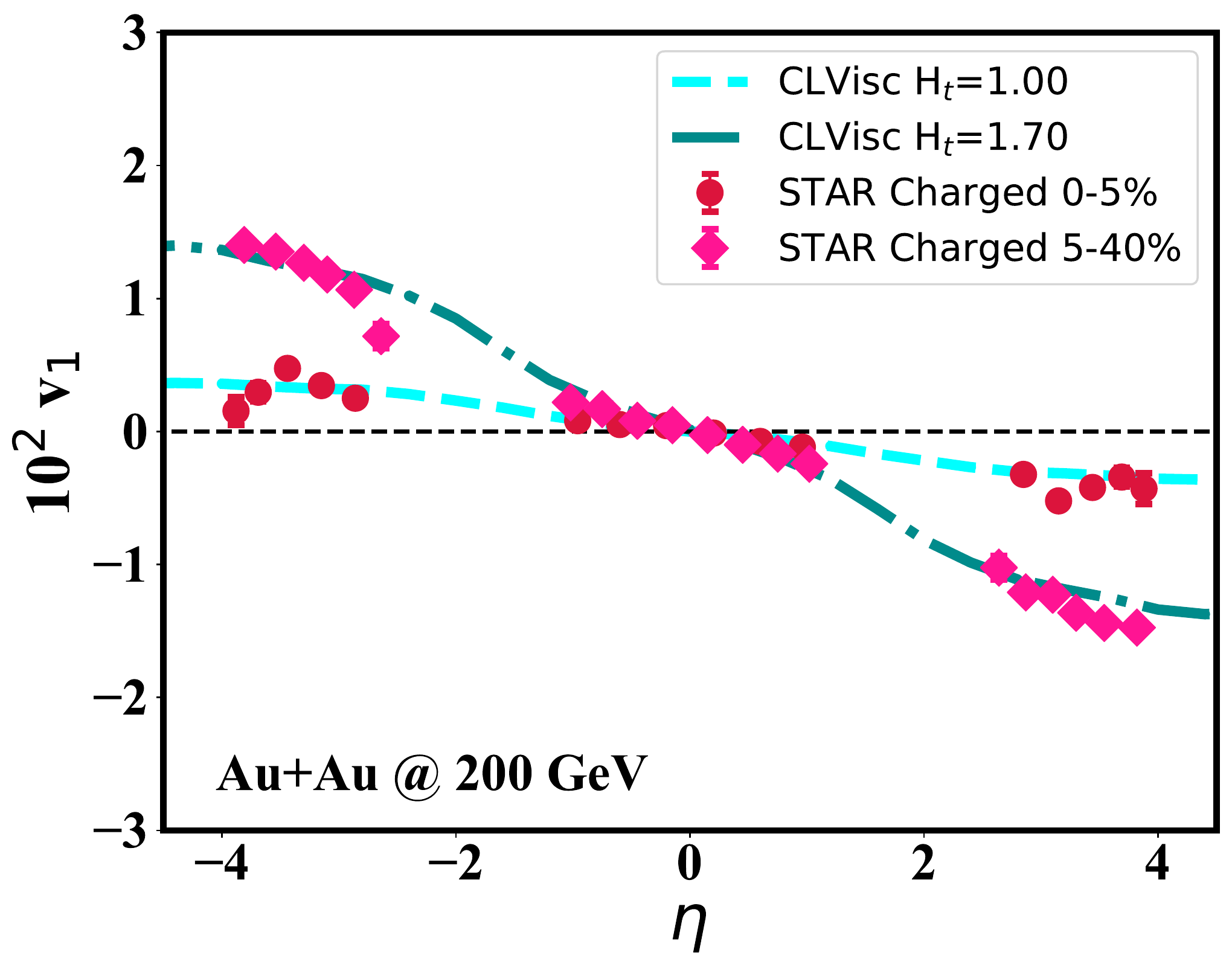}~ \\
\includegraphics[width=0.85\linewidth]{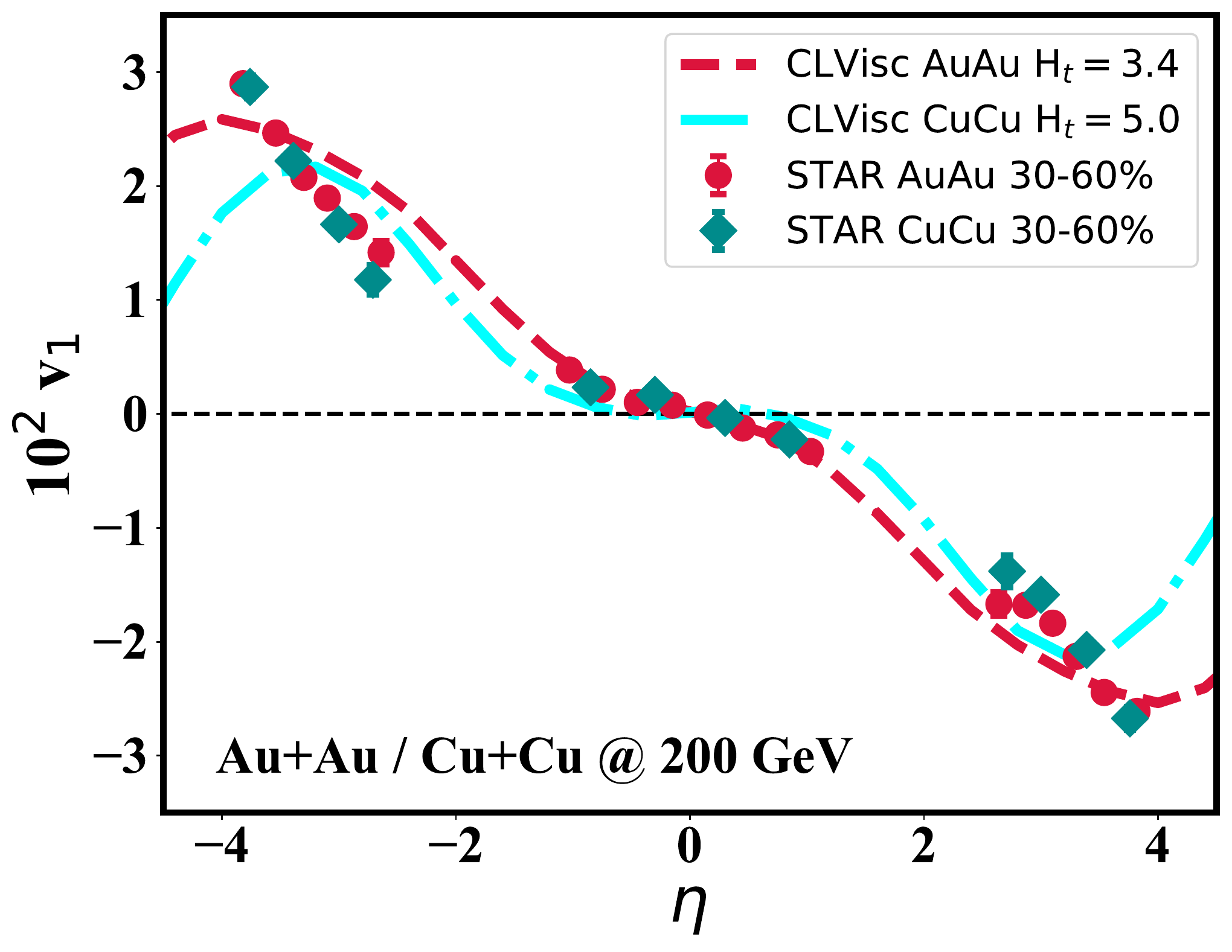}~
\end{center}
\caption{(Color online) Directed flow coefficients $v_{1}(\eta)$ of charged particle versus pseudorapidity from CLVsic (colored curves)
for Au+Au and Cu+Cu $\sqrt{s_{NN}}=200$ GeV collisions in comparison
to the experiment data from STAR Collaboration at RHIC energy(solid points)~\cite{Abelev:2008jga}.
(top panel) Results for centrality class 0-5\%, 5\%-40\%. (bottom pane) Results for centrality class 30\%-60\%.}
\label{f:rhicdf}
\end{figure}

However, a non-zero directed flow $v_{1}$ can be generated in the hydrodynamic simulation with the modified initial conditions.
Figure~\ref{f:rhicdf} shows the result for the directed flow coefficient $v_{1}(\eta)$ of charged particles emitted after a hydrodynamic evolution.
The dashed (-dotted) curves are the results for Au + Au and Cu + Cu $\sqrt{s_{NN}}=200$ GeV collisions. One finds
the experimental data in the centrality classes 0\%-5\% and 5\%-40\% are reproduced well in a large pseudorapidity region.
{\color{black} For the peripheral collisions (in the centrality bin 30\%-60\%), our model overestimates the directed flow coefficient $v_{1}$ at large rapidity (fragmentation region).
The directed flow in such a region maybe has different origins, such as the baryon stopping effect and fluctuations~\cite{Petersen:2006vm,Bozek:2010bi}.}

\subsection{Pb+Pb $\sqrt{s_{NN}}$ = 2.76 TeV and 5.02 TeV collisions}

\begin{figure}[!htb]
\begin{center}
\includegraphics[width=0.8\linewidth]{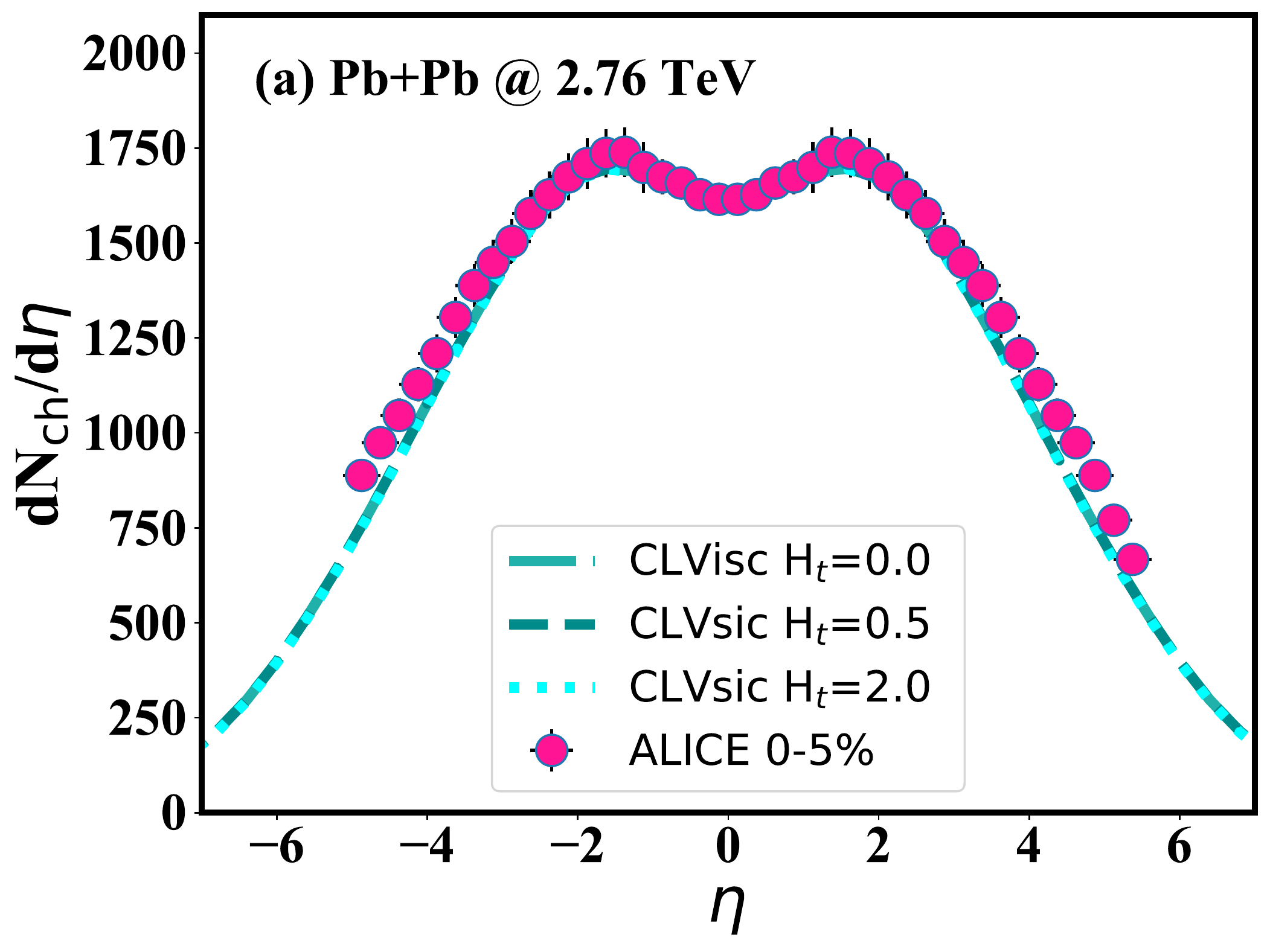}~ \\
\includegraphics[width=0.8\linewidth]{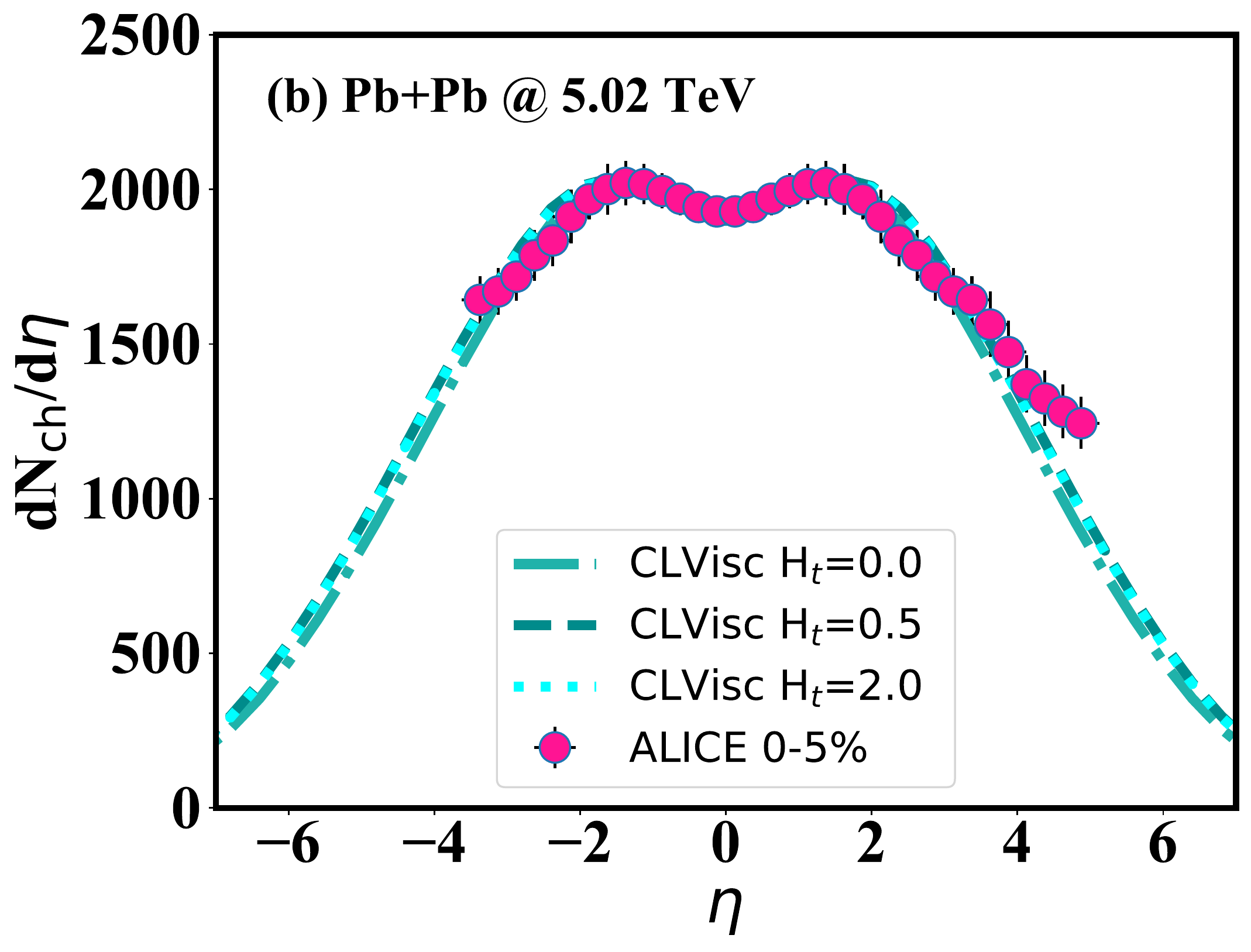}~
\end{center}
\caption{(Color online) The charged particle pseudorapidity distribution from CLVsic (colored curves) for Pb+Pb $\sqrt{s_{NN}}$ = 2.76 TeV~\cite{Adam:2015kda} and $\sqrt{s_{NN}}$ = 5.02 TeV~\cite{Adam:2016ddh} collisions at the LHC energy in comparison to the experiment data from ALICE Collaboration (solid points).}
\label{f:lhcdn}
\end{figure}

Figure~\ref{f:lhcdn} shows only the most central pseudorapidity distribution (0-5\%) for charged particles from the Ideal-CLVisc is
in comparison with experimental data from the
ALICE Collaboration~\cite{Adam:2015kda,Adam:2016ddh}.
The hydrodynamic simulation with different longitudinal tilted initial condition ($H_{t}$ = 0.0, 0.5, 2.0)
gives almost the same charged multiplicity distribution dN/d$\eta$ for the most central collisions.

\begin{figure}[!htb]
\begin{center}
\includegraphics[width=0.8\linewidth]{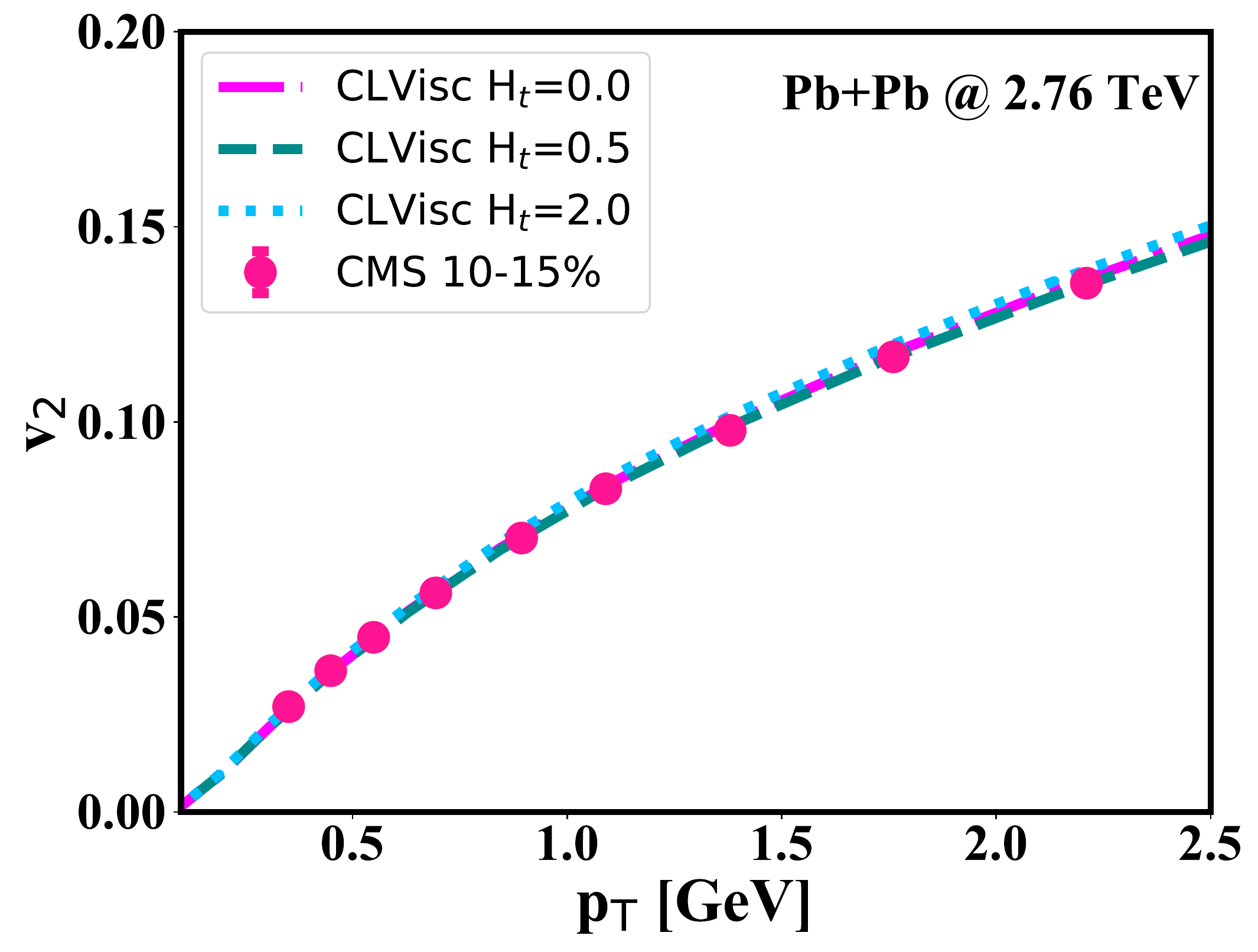}~\\
\includegraphics[width=0.8\linewidth]{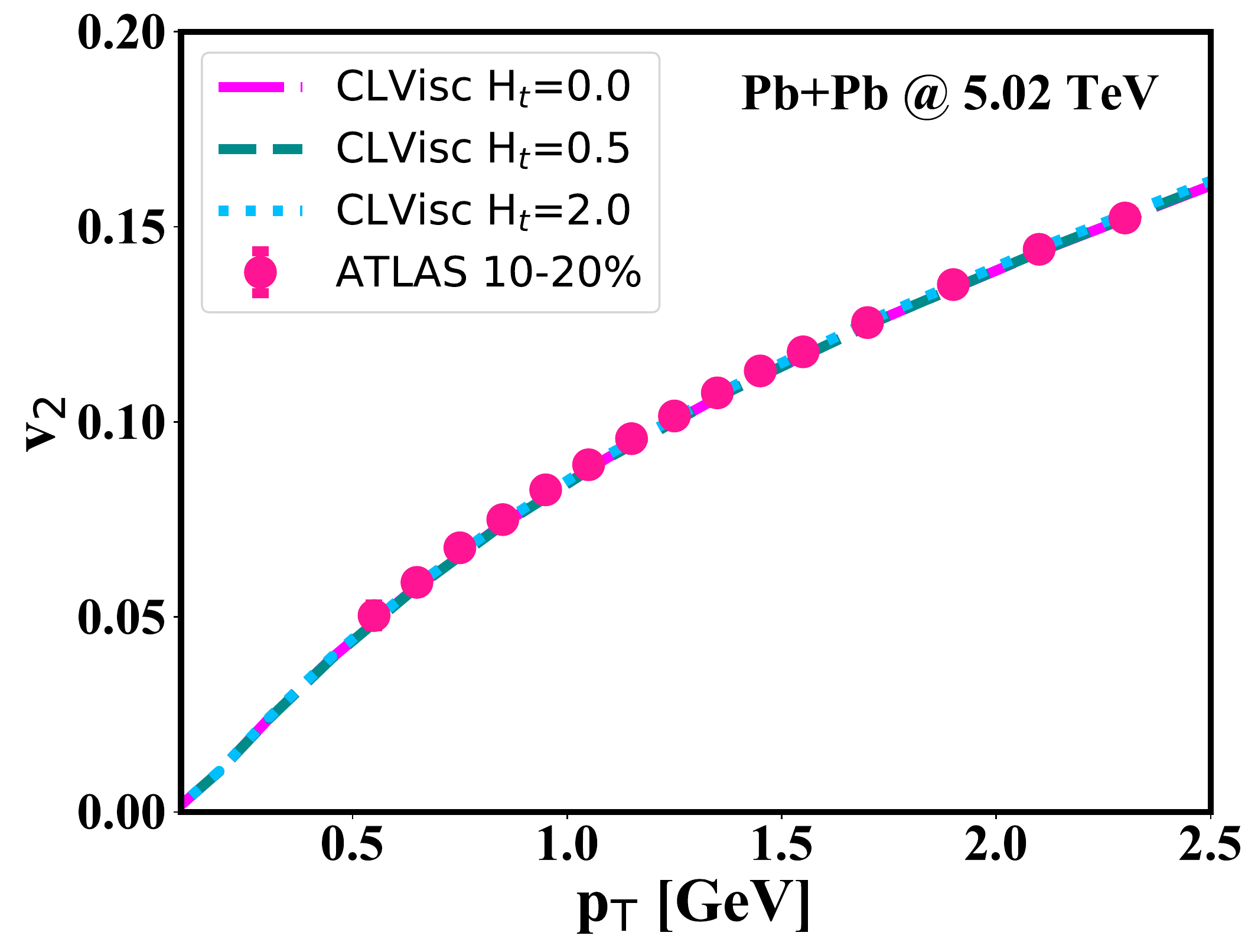}~
\end{center}
\caption{(Color online) Elliptic flow coefficients $v_{2}(p_{T})$ for Pb + Pb collisions at the LHC energy.
Top panel: result in 10\%-15\% Pb + Pb collisions at $\sqrt{s_{NN}}=2.76$ TeV,
the experimental data comes form CMS Collaboration~\cite{Chatrchyan:2012ta}.
Bottom panel: results for Pb+Pb $\sqrt{s_{NN}}=5.02$ TeV collisions for the centrality class 10\%-20\%,
the experimental data comes from ATLAS Collaboration~\cite{Aaboud:2018ves}.}
\label{f:lhcv2}
\end{figure}

In Figure~\ref{f:lhcv2} we plot the elliptic flow $v_{2}(p_{T})$ only for Pb + Pb $\sqrt{s_{NN}} = 2.76$ TeV collisions in the centrality bin 10\%-15\% (top panel)
and Pb+Pb $\sqrt{s_{NN}}~=~5.02$ TeV collision (bottom panel) in the centrality bin 10\%-20\%.
We find that the elliptic flow $v_{2}(p_{T})$  is indistinguishable for different $H_{t}$ (dashed and dashed-dotted curves).
The experimental data are from the CMS Collaboration~\cite{Chatrchyan:2012ta}
and ATLAS Collaboration~\cite{Aaboud:2018ves}.

\begin{figure}[!ht]
\begin{center}
\includegraphics[width=0.8\linewidth]{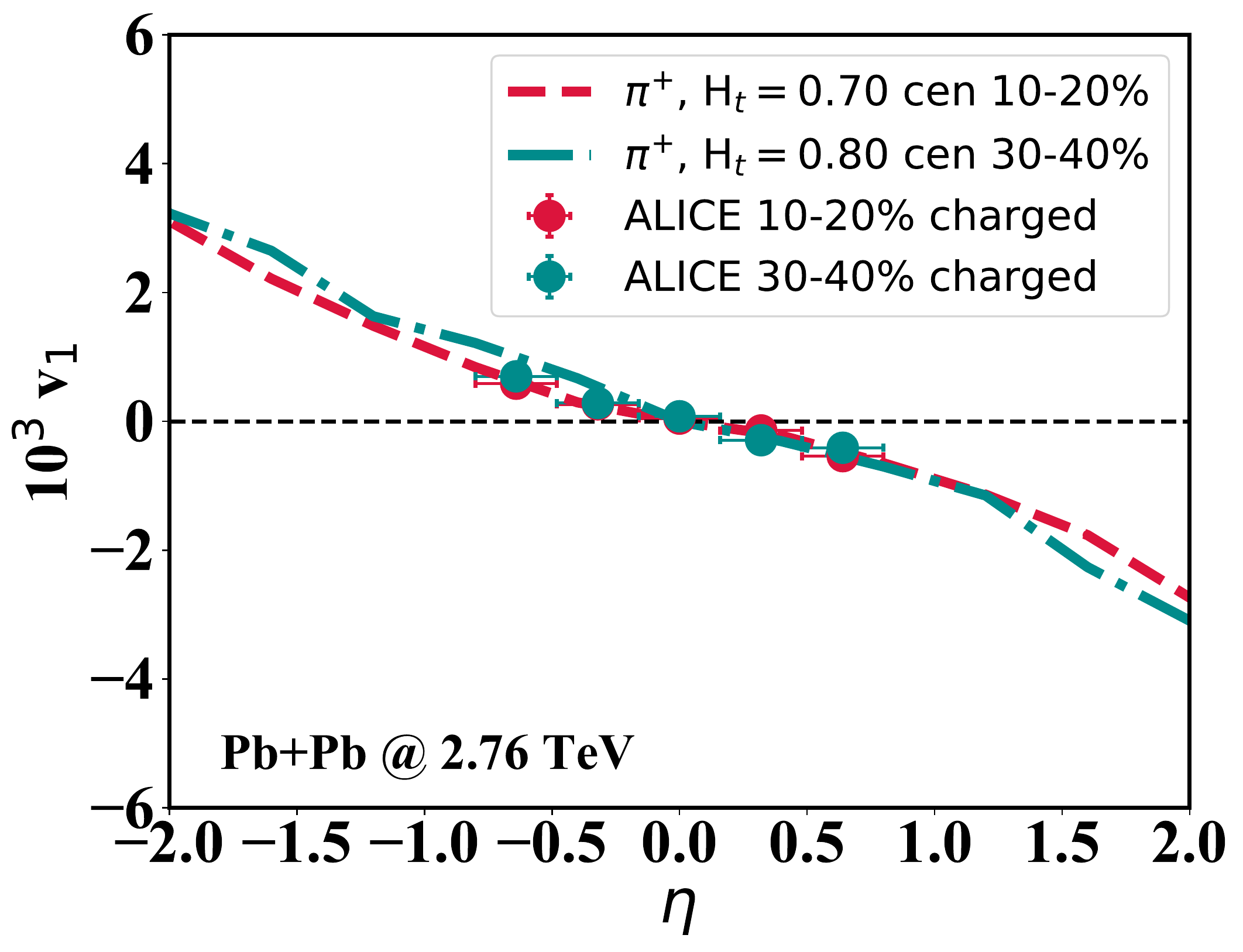}~ \\
\includegraphics[width=0.8\linewidth]{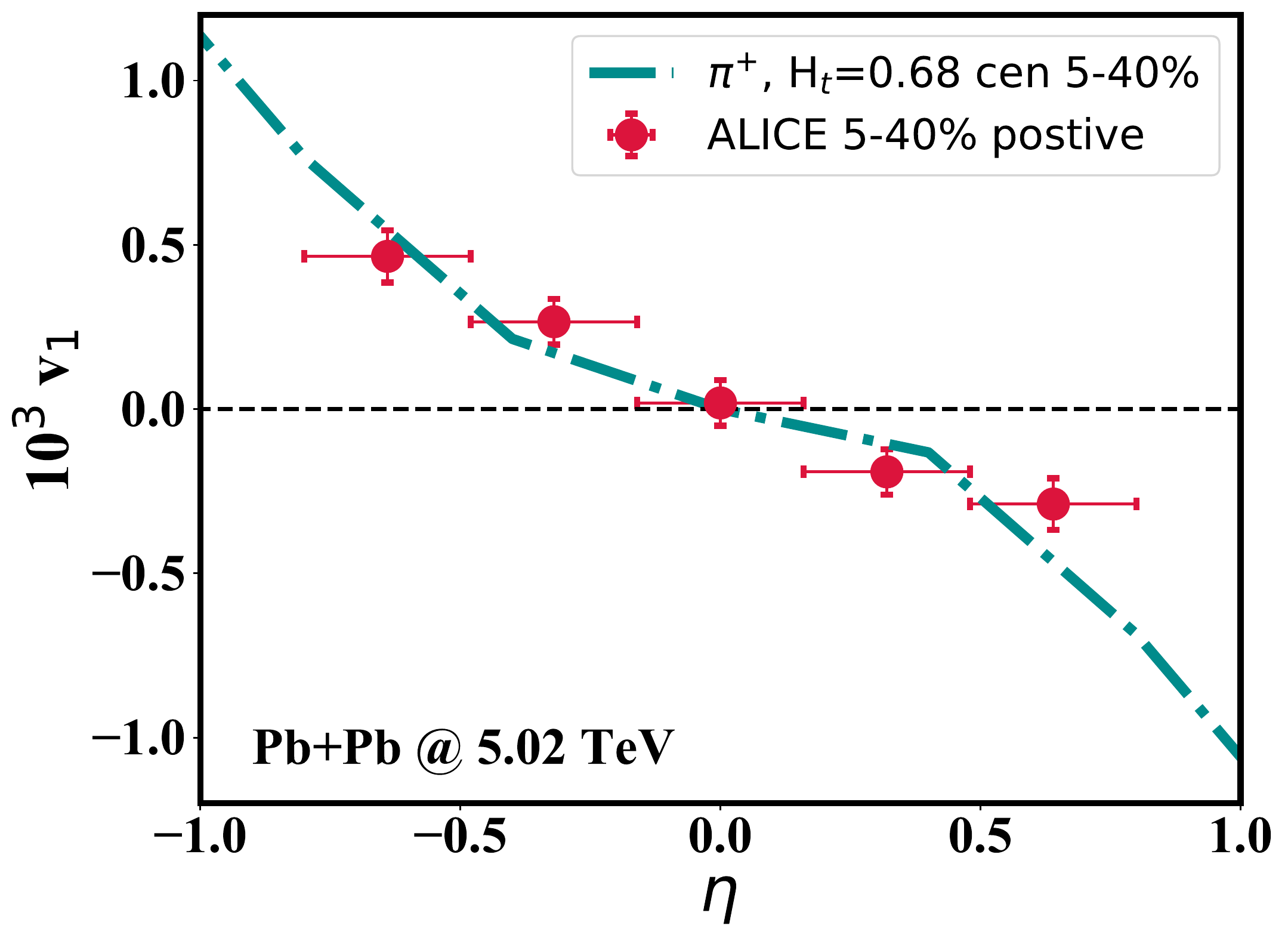}~
\end{center}
\caption{(Color online) Directed flow coefficients $v_{1}(\eta)$ of $\pi^{+}$ versus
pseudorapidity from CLVsic (colored curves) for Pb+Pb collisions at the LHC energy in comparison
with the experiment data from ALICE Collaboration (solid points)~\cite{Abelev:2013cva,Acharya:2019ijj}.
(top panel) Results for Pb+Pb $\sqrt{s_{NN}}$ = 2.76 TeV collision at centrality class 10\%-20\%, 30\%-40\%.
(bottom panel) Results for Pb+Pb $\sqrt{s_{NN}}$ = 5.02 TeV collision at centrality class 5\%-40\%.}
\label{f:lhcv1}
\end{figure}

In Figure~\ref{f:lhcv1} we plot the pseudorapidity dependence of the directed flow $v_{1}(\eta)$ for pions ($\pi^{+}$).
Figure~\ref{f:lhcv1} (top panel) shows the results for Pb + Pb $\sqrt{s_{NN}}=2.76$ TeV in the centrality classes 10\%-20\% and 30\%-40\%.
We see that hydrodynamic simulation with modified initial conditions reproduce the experimental data from ALICE Collaboration~\cite{Abelev:2013cva}
in the centrality bin 10\%-20\% with tilted parameter $H_{t}$ = 0.7 and 30\%-40\% with H$_{t}$ = 0.8.

Figure~\ref{f:lhcv1} (bottom panel) shows the result for Pb + Pb $\sqrt{s_{NN}}=5.02$ TeV in the centrality class 5\%-40\%.
We see that the ALICE Collaboration~\cite{Acharya:2019ijj} points for the centrality class 5\%-40\% are for positive charged particles,
and results for the pion ($\pi^{+}$) directed flow from CLVisc simulation agree with the experimental data to a reasonable level.
This is because (a) a distinction for the whole positive and negative particle yields needs a proper theory, which has not been developed totally in the current stage;
(b) $\pi^{+}$ plays a vital role for the yield of the total positive charge particles and mainly comes from the hydrodynamic evolution~\cite{Adam:2015kda,Adam:2016ddh}.

\begin{figure}[!ht]
\begin{center}
\includegraphics[width=0.8\linewidth]{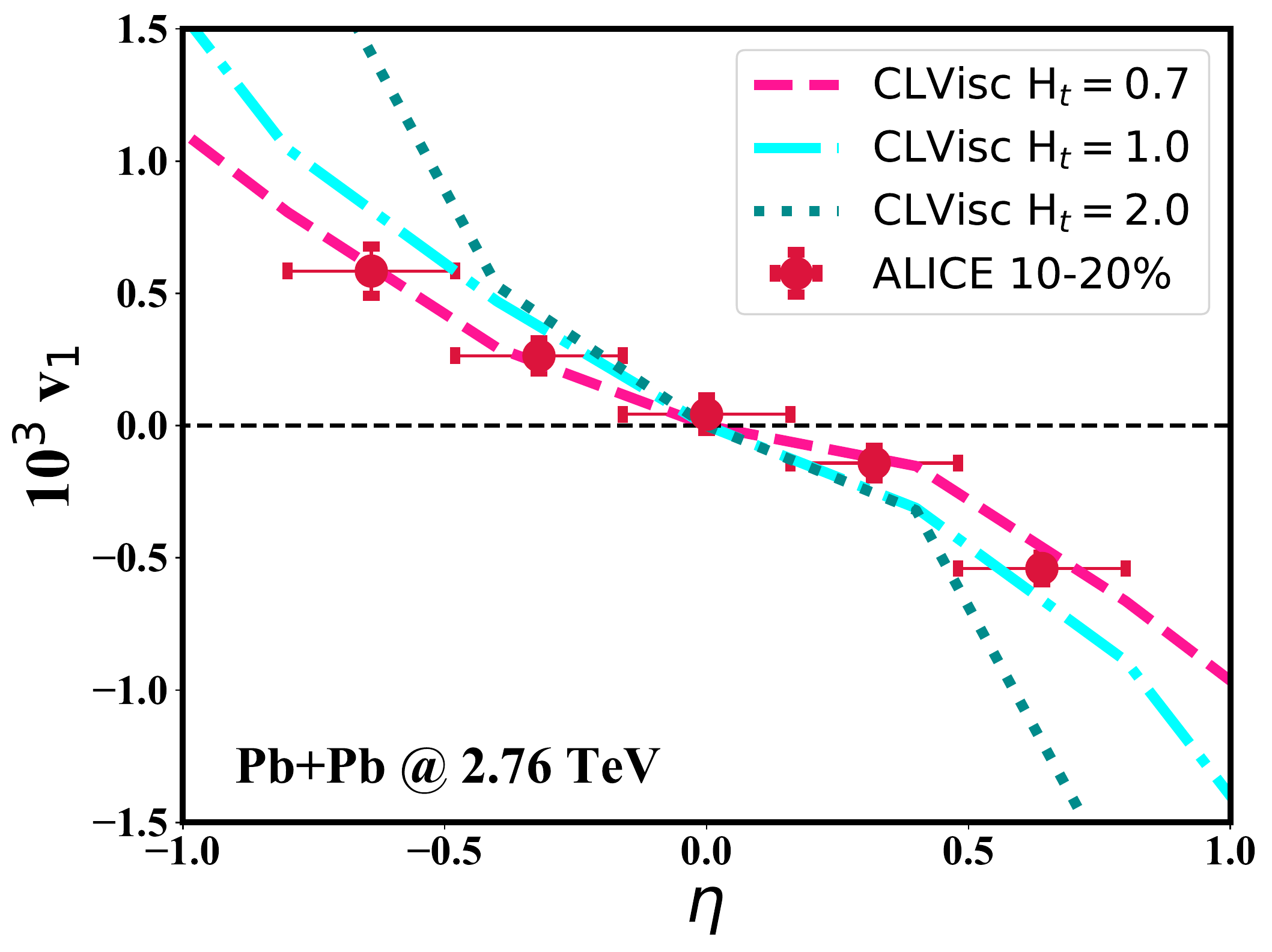}
\end{center}
\caption{(Color online)Directed flow coefficients $v_{1}(\eta)$ of charged particles versus pseudorapidity from CLVsic (colored curves)
for Pb+Pb collisions at the LHC energy in comparison
to the experiment data from ALICE Collaboration with different model parameter H$_{t}$.}
\label{f:v1H}
\end{figure}

To get information that how the longitudinal tilted structure introduced in Eq.~(\ref{eq:mwneta1}) is related to the final azimuthal asymmetry measured
by the directed flow,
we plot the $H_t$ dependence of the directed flow of charged particles in Figure~\ref{f:v1H} for Pb + Pb $\sqrt{s_{NN}}$ = 2.76 TeV collisions in the centrality bin 10\%-20\%.
We find a larger value of $H_{t}$ corresponds to a more longitudinal tilted fireball and leads to a larger value of the directed flow coefficient at large rapidity.
The value of $H_{t}$ extracted from the STAR and ALICE data show that
(a) the initial spatial pressure gradient asymmetry in the transverse plane of bulk medium at RHIC energy is larger than at the LHC energy at the initial proper time;
(b) the larger the impact parameter $b$, the larger the magnitude asymmetry of initial pressure gradient in the transverse plane at the initial stage.

\section{Summary}
\label{v1section4}

In this work, {\color{black} following the previous work Refs.~\cite{Bozek:2011ua,Bozek:2010bi,Inghirami:2019mkc,Beraudo:2021ont},
an alternative parametrization to construct longitudinal tilted initial condition based on the Glauber model is presented.}
A data-driven phenomenological rapidity-dependent weight function [Eq. (\ref{eq:mwneta1})]
is obtained to generate the initial tilting longitudinal expansion of the strongly-coupled QCD matter,
and the results show that such a tilting source leads to a magnitude asymmetry of the pressure gradient along the $x$-direction
and longitudinal $\eta_{s}$ direction~\cite{Bozek:2010bi}.
The existence of such asymmetries push the light quarks and gluons toward the $x>0$ (and $x<0$) direction at different rapidities,
which gives a nonzero directed flow coefficient.

The pseudorapidity distribution $dN/d\eta$ and the elliptic flow coefficient $v_{2}(p_{T})$ of charged particles are presented for Cu + Cu, Au + Au and Pb + Pb collisions for serval longitudinal tilted parameters $H_{t}$.
The longitudinal tilted structure in hot-QCD matter affect both the multiplicity density distribution and elliptic flow coefficient weakly and almost negligible, consistent with
previous studies~\cite{Bozek:2011ua,Bozek:2010bi}.

{\color{black}Our modified initial conditions with the ideal CLVisc hydrodynamic simulation is used to fit the directed flow coefficient $v_{1}(\eta)$ of charged particle and pions 
measured by the STAR Collaboration and the ALICE Collaboration.} We find that the directed flow coefficient is generated at a very early stage in the evolution
and is given by the initial tilted source. The directed flow coefficient is decreased with the center-of-mass energy $\sqrt{s_{NN}}$, the reasons are
(a) with increasing range of longitudinal rapidity the imbalance between the thickness function $T_{1}$ and $T_{2}$ goes down;
(b) the contribution of binary collisions to the Glauber model is believed to increase with the $\sqrt{s_{NN}}$~\cite{Bozek:2011ua,Inghirami:2019mkc}.

We also remark that, besides the initial fireball spatial asymmetry contributing to the directed flow,
the following aspects are also important and can be extended to future work.

(1) In real heavy-ion collisions, the fluctuations of the initial energy density also contribute to the local pressure gradient asymmetry,
\eg T$_{\textrm{R}}$ENTo-3D initial condition model~\cite{Ke:2016jrd}.
It is possible that the dynamical fluctuations or the initial flow in the $z$ direction
can generate a large pressure gradient asymmetry for both angle and magnitude~\cite{Ke:2016jrd,Bozek:2010bi,Chen:2019qzx,Shen:2020jwv,Oliva:2020doe}.

(2) {\color{black}During the hydrodynamic expansion, the viscosity corrections reduce the longitudinal pressure and increase the transverse pressure.
As a result, a smaller directed flow coefficient is observed after a viscous hydrodynamic simulation.
This means that the ideal hydrodynamics is a suitable tool for the study of the presence of directed flow in heavy ion collisions~\cite{Bozek:2011ua}.}
The expansion of the strongly coupled matter is affected by the shear viscosity and bulk viscosity in both transverse plane and longitudinal direction.
The observables such as the elliptic flow and $p_{T}$ spectra are more sensitive to the viscosity
effect according to viscous hydrodynamics calculations~\cite{Pang:2018zzo,Hirano:2002ds,Hirano:2005xf,Song:2007fn,Romatschke:2007mq}.
Moreover, the $v_{2}(\eta)$ coefficient is sightly overestimated in our model as the lack of the shear viscosity effect~\cite{Pang:2018zzo,Bozek:2011ua}.
Because $v_{2}(\eta)$ places severe tight constraints to the initial shape of the fireball away from midrapidity,
and thus to the parametrization of the initial state~\cite{Pang:2018zzo,Hirano:2002ds,Hirano:2005xf}.
In the future study, viscous correction will be included and the $v_{2}(\eta)$ coefficient will be used to constrain our initial condition.

(3) For non-central heavy ion collisions, an extremely strong magnetic-field is
created by the colliding charged beams moving at relativistic speed (almost $10^{16}-10^{20}$ Gauss)~\cite{Li:2016tel,Zhong:2014sua,Pu:2016ayh,She:2019wdt}.
Due to the expansion along the beam axis, the Lorentz force is directed along the negative-$x$ direction in the forward rapidity region for positively charged quarks,
which generates a directed transverse flow.
In addition to the above Hall effect, the time dependence of the magnetic-field generates an electric field due to the Faraday effect.
The induced Faraday current provides a no-zero finite drift velocity in the transverse plane due to the magnetic field.
{\color{black} Lots of work has investigated the contribution of the combination of the above two effects on the directed flow coefficient,
such as the MHD~\cite{Inghirami:2019mkc} and decoupled hydro-magnetic frame~\cite{Gursoy:2014aka,Gursoy:2018yai}.
And one has found that the contribution of the magnetic field effect on the soft hadron directed flow coefficient is less than $5\times 10^{-4}$ at large rapidity,
which is almost 80 - 100 times smaller than the contribution from the tilted initial conditions~\cite{Inghirami:2019mkc}.}

(4) Taking into account the asymmetry between forward and backward moving participants, the noncentral
heavy ion collisions produce not only strong angular momentum, strong magnetic-field but also global and local
vorticity and hyperon polarization~\cite{Pang:2016igs,Oliva:2020doe}.

(5) The directed flow of heavy hadrons could be a great probe to investigate the initial pressure gradient asymmetry of bulk matter~\cite{Chen:2019qzx,Das:2016cwd}.
Vice versa, such kind of tilted medium could be the background of heavy quark propagation, and
its effect on open heavy flavor production $v_{1}$ and $R_{AA}$ might be interesting problems~\cite{Cao:2016gvr,Prado:2019ste,Xing:2019xae,Li:2020kax,Beraudo:2021ont}.
Furthermore, $D^{0}$ directed flow found at both STAR and ALICE still contains large statistical uncertainly and systemic uncertainly,
one suggests that hydrodynamic + transport model together may put
more constraints on the directed flow for better understanding of the initial stage in heavy ion collisions~\cite{Dubla:2020bdz}.

(6) {\color{black}We also need to take into account the hadronic cascade.} For a proper comparison with more experimental data,
one should include such interactions with the addition of hadronic transport models such as UrQMD~\cite{Bass:1998ca,Zhao:2021vmu} and SMASH~\cite{Petersen:2018jag,Wu:2021fjf}.

(7) Recently, a large number of studies used the parametrizations for the longitudinal structure of the fireball in Ref.~\cite{Bozek:2010bi} to investigate
the directed flow coefficient of heavy meson~\cite{Chatterjee:2018lsx,Beraudo:2021ont,Chatterjee:2017ahy,Oliva:2020doe}.
Collision geometry-based 3D initial conditions with hydrodynamic simulations from the group of \emph{C. Shen} et al.~\cite{Shen:2020jwv,Ryu:2021lnx} also described
the directed flow coefficient of $\pi^{+}$ at RHIC energies well.
It will be interesting to see which one of the models can lead to the reproduction of the experimental data with the least amount of tweaking of the parameters.

These important aspects will be studied in the future.

\begin{acknowledgements}
The authors thank Xin-Nian Wang for helpful comments and providing the GPU computing platform at the initial stage of this study.
Z-F. Jiang would like to thank Xiangyu Wu, ShanShan Cao, Chun Shen, Chi Ding, Zhong Yang and Long-Gang Pang for helpful discussion.
{\color{black} The authors would like to thank the anonymous reviewers for their helpful remarks.}
This research was supported by the NSFC of China with Project No.~11935007, Hubei Provincial Natural Science Foundation of China, the Education Department of Hubei Province of China with Young Talents Project No.~Q20212703 and
the Xiaogan Natural Science Foundation of China under Grant No.XGKJ2021010016. Computational resources have been provided by the Center of Scientific Computing
at the Department of Physics and Electronic-Information Engineering, Hubei Engineering University.
Computational resources have been provided by the Center of Scientific Computing
at the Department of Physics and Electronic-Information Engineering, Hubei Engineering University.
The numerical simulations have been preformed partly at the GPU cluster in the Nuclear Science Computing Center at Central China Normal University(NSC$^3$).
\end{acknowledgements}

\bibliographystyle{unsrt}
\bibliography{v1ref}

\end{document}